\documentclass[aps,pre,showpacs,nofootinbib,superscriptaddress,reprint,one,notitlepage]{revtex4}

\usepackage{amsfonts}
\usepackage{amsmath}
\usepackage{amssymb}
\usepackage{graphicx}
\usepackage{color}
\usepackage{mathrsfs}
\usepackage{bbm}
\usepackage{subfigure}
\usepackage{enumerate}
\usepackage{xcolor}
\usepackage{float}
\usepackage{times,txfonts}
\usepackage{ulem}
\usepackage{siunitx}
\usepackage{bbold}
\usepackage{sidecap}
\usepackage{mathtools}
\newcommand{\ignore}[1]{}
\newcommand{\nobibentry}[1]{{\let\nocite\ignore\bibentry{#1}}}

\newcommand*{\rom}[1]{\expandafter\@slowromancap\romannumeral #1@}
\makeatother

\DeclareMathOperator\arctanh{arctanh}

\begin{document}

\title{Thermodynamics of quantum feedback cooling}

\author{Pietro Liuzzo-Scorpo}
\affiliation{School of Mathematical Sciences, The University of Nottingham, University Park, Nottingham NG7 2RD, United Kingdom}
\author{Luis A. Correa}
\affiliation{Unitat de F\'isica Te\`orica: Informaci\'o i Fen\`omens Qu\`antics, Departament de F\'isica, Universitat Aut\`onoma de Barcelona, 08193 Bellaterra (Barcelona), Spain}
\author{Rebecca Schmidt}
\affiliation{Centre for Quantum Engineering and Centre of Excellence in Computational
	Nanoscience, Department of Applied Physics, Aalto University School of Science, P.O.Box 11100, 00076 Aalto, Finland}
\affiliation{Turku Centre for Quantum Physics, Department of Physics and Astronomy, University of Turku, FI-20014, Turun Yliopisto, Finland}
\author{Gerardo Adesso}
\affiliation{School of Mathematical Sciences, The University of Nottingham, University Park, Nottingham NG7 2RD, United Kingdom}
\email{gerardo.adesso@nottingham.ac.uk}

\begin{abstract}
	The ability to initialize quantum registers in pure states lies at the core of many applications of quantum technologies, from sensing to quantum information processing and computation. In this paper, we tackle the problem of increasing the polarization bias of an ensemble of two-level register spins by means of joint coherent manipulations, involving a second ensemble of ancillary spins and energy dissipation into an external heat bath. We formulate this spin refrigeration protocol, akin to algorithmic cooling, in the general language of quantum feedback control, and identify the relevant thermodynamic variables involved. Our analysis is two-fold: on the one hand, we assess the optimality of the protocol by means of suitable figures of merit, accounting for both its work cost and effectiveness; on the other hand, we characterise the nature of correlations built up between the register and the ancilla. In particular, we observe that neither the amount of classical correlations nor the quantum entanglement seem to be key ingredients fuelling our spin refrigeration protocol. We report instead that a more general indicator of quantumness beyond entanglement, the so-called quantum discord, is closely related to the cooling performance.
\end{abstract}


\maketitle

\section{Introduction}\label{sec:intro}

Achieving low enough temperatures is an essential prerequisite to bring physical systems into the quantum domain. For instance, a thermal cloud of atoms may be coerced into a Bose--Einstein condensate \cite{anglin2002bose} once chilled below the micro-Kelvin range by means of laser and evaporative \linebreak cooling \cite{phillips1998laser,masuhara1988evaporative}. Similarly, quantum behaviour can be observed in mesoscopic objects by exploiting active cooling strategies, which outperform conventional passive refrigeration methods \cite{hopkins2003feedback,kleckner2006sub,poggio2007feedback}. Hence, the exploitation of quantum effects in most technological applications relies heavily on our ability to generate ultracold systems on cue.

Over the past few years, there has been an intense activity on \textit{quantum thermodynamics} \cite{1310.0683v1,gelbwaser2015thermodynamics} and, specifically, on the study of nanoscale cooling cycles \cite{e15062100,Koski2015,Kutvonen2015}. Various models of quantum refrigerators have been put forward and characterized \cite{PhysRevE.64.056130,gelbwaser2014heat,PhysRevE.89.042128}: although the focus has often been placed on the fundamental problem of the emergence of the thermodynamic laws from quantum theory \cite{rezek2009quantum,PhysRevLett.109.090601,PhysRevE.85.061126}, more practical issues, like cooling performance optimization \cite{PhysRevE.81.051129,Correa2013,PhysRevE.90.062124} or cycle diagnosis in the search for friction, heat leaks and internal dissipation \cite{kosloff2010optimal,correa2015internal,PhysRevE.73.025107}, have been addressed.

Some of the recent landmark achievements in experimental physics
have provided us with a complete ``quantum toolbox'', facilitating the controlled manipulation of individual quantum systems. This raises the expectations that practical nanoscale heat devices may soon become commonplace in technological applications. Indeed, experimental proposals exist for quantum refrigerators \cite{0295-5075_97_4_40003,PhysRevLett.110.256801}, and promising new cooling methods have already been demonstrated \cite{PhysRevLett.110.157601,PhysRevA.91.023431}. These include feedback cooling \cite{hopkins2003feedback,steck2004quantum,Koski2015,Kutvonen2015} of individual quantum systems \cite{bushev2006feedback} and even of nano- and micro-mechanical resonators \cite{kleckner2006sub,poggio2007feedback}.

Another big open question in quantum thermodynamics is whether or not quantum signatures may be actively exploited in nanoscale heat cycles. It is known, for instance, that the discreteness of the energy spectra of quantum devices allows for departures from the usual thermodynamic behaviour when non-equilibrium environments are considered \cite{1303.6558v1,Correa2014,PhysRevLett.112.030602,alicki2015non}. Furthermore, distinct non-classical signatures stemming from quantum coherence may be observed directly in quantum heat \linebreak engines \cite{niedenzu,PhysRevX.5.031044}. Nonetheless, in most respects, nanoscale heat devices closely resemble their macroscopic counterparts \cite{alicki1979engine,kosloff1984quantum}. Importantly, instances showing some indicator of quantumness being actively {utilized} for better-than-classical energy conversion are still essentially missing.

In this paper, we will ask precisely this type of question regarding a quantum feedback cooling loop. Namely, is the quantum share of the correlations being established between the system of interest and the controller a resource for energy-efficient feedback cooling?

To address it, we shall consider the algorithmic cooling \cite{boykin2002algorithmic,fernandez2004algorithmic} of spins in nuclear magnetic resonance (NMR) setups. Concisely, algorithmic cooling aims at increasing the polarization bias of an ensemble of spins, hereafter labelled as ``registers''. To that end, a second ensemble of spins (``ancillas'') with a larger polarization bias is available. A suitable series of joint quantum gates is then applied in order to reversibly dump a fraction of the registers' entropy into the ancillas. As a final step, the ancillary spins are dissipatively reset back to their initial state, thus disposing of their excess entropy into the surroundings. Since the relaxation time of the registers is assumed to be much longer than that of the ancillas, the latter are reset, whilst the former remain essentially unchanged.

The usage of the surroundings as an external entropy sink is crucial for the scalability of the spin cooling protocol \cite{fernandez2004algorithmic}. Indeed, reversible manipulations preserve the total entropy of a closed system, which sets an ultimate limitation (``Shannon's bound'') on the achievable spin cooling. However,~after the reset of the ancillas, the cooling algorithm may be iterated to further reduce the entropy of the registers. This cooling technique has also been demonstrated experimentally \cite{baugh2005experimental,ryan2008spin}.

We shall work in the framework of coherent feedback control \cite{Lloyd2000,habib2002quantum}, splitting the global unitary manipulation of registers and ancillas into two, as if the transformation were to be performed in two distinct steps: namely, the ``(pre)measurement'' and the ``feedback'' itself. The purpose of the measurement step should be to build correlations between register and ancillary spins, allowing the controller to acquire useful information. The feedback unitary should then exploit that information to reduce the entropy of the registers as much as possible. One of the aspects we wish to understand is the specific role of the quantum share of the correlations in the operation of the cooling cycle.

In the first place, we will study the energy balance throughout one iteration of the protocol, identifying the relevant quantities standing for ``work'', ``heat'' and ``entropy reduction rate''. We~shall then introduce suitable figures of merit assessing the \textit{energy efficiency} and the \textit{effectiveness} of the feedback loop. These will allow us to identify thermodynamically optimal working points. Finally,~we will attempt to establish connections between performance optimization and the correlations built up between registers and ancillas: we will find that, unlike the classical correlations, the quantum share thereof (measured by the quantum discord \cite{olliver20011,henderson20011}) relates closely to the performance of the cycle. In particular, the maximization of discord is compatible with the thermodynamic optimization of the protocol, and a threshold value of the quantumness of correlations exists, which enables effective cooling of the registers. It will also be shown that quantum entanglement between registers and ancillas, though widely present, is not a key ingredient in the protocol. Ultimately, these results may help to clarify the intriguing role of quantum correlations in the thermodynamics of information \cite{parrondo2015thermodynamics} and pave the way towards their active exploitation in practical tasks \cite{sagawa2008second,park2013heat}.

This paper is structured as follows: In Section~\ref{sec:protocol}, coherent feedback control is briefly introduced and our specific cooling cycle described in detail. The energetics of a single iteration of our protocol is addressed in Section~\ref{sec:thermodynamics}, where we also introduce objective functions to gauge its cooling performance. In Section~\ref{sec:information}, entanglement, discord and quantum mutual information between register and ancillary spins are evaluated explicitly. Finally, in Section~\ref{sec:conclusions}, we summarize and draw our conclusions.

\section{Feedback Cooling Algorithm}
\label{sec:protocol}

\vspace{-6pt}
\subsection{Coherent Feedback Control}

Algorithmic cooling may be thought of as an application of \textit{feedback control} \cite{Dong2010, Wise2010}. In general, preparing a system in a desired configuration is one of the fundamental tasks of control theory.
\linebreak This can be achieved by introducing a work source to the setting, whose interactions with the system of interest are suitably engineered to achieve the goal. Nonetheless, in practical situations, a direct access to the system might not be possible or it may accompanied by irrepressible disturbances. \linebreak One thus typically introduces a second (auxiliary) system into the problem. The composite formed by the work source plus the auxiliary system can then be referred to as the \textit{controller} \cite{PhysRevA.62.012105}. While in an ``open loop'' control protocol the controller does not acquire any information about the system during the control process (\textit{i.e.}, the interaction is unidirectional), in ``closed-loop'' control (or feedback control) the controller gains information about the state of the system that conditions its subsequent actions~\cite{habib2002quantum,Touch2004}.

In our case, the system of interest is quantum, which raises the measurement problem \cite{Yama2014}, \textit{i.e}., how the information is actually gained. On the one hand, it may be obtained by performing an explicit measurement on the system \cite{Wise1993,sagawa2008second}, which entails the erasure of quantum coherence in some pre-selected basis. In such case, we would be performing a classical feedback \cite{Lloyd2000}. Alternatively, we may as well correlate system and auxiliary, without making the measurement explicit, and then close the feedback loop coherently. This is the situation that we shall consider here: a coherent \textit{quantum} feedback control protocol \cite{Lloyd2000,Gough2009,park2013heat}.

\subsection{Stages of the Feedback Cooling Algorithm}

In all that follows, both our quantum registers (the system $ \mathsf{S} $) and the ancillas (the auxiliary $ \mathsf{A} $) will be two-level spins. We shall assume that they are initially uncorrelated and in thermal equilibrium with the surroundings, which act as a heat bath at temperature $ T $. We denote the polarization bias of the registers and the ancillas, \textit{i.e}.,~the difference between their ground and excited-state populations, by $ \epsilon_\mathsf{S} $ and $ \epsilon_\mathsf{A} $, respectively. As already mentioned, in order to reduce the entropy of the spin ensemble $ \mathsf{S} $, two conditions have to be met: (i) $ \epsilon_\mathsf{A}>\epsilon_\mathsf{S} $, \textit{i.e}.,~the ancillas must be more polarized than the registers at the initialization stage; and (ii) the relaxation time of $ \mathsf{A} $ must be much shorter than that of $ \mathsf{S} $.

\begin{figure}[H]
	\centering
	\includegraphics[scale=0.35]{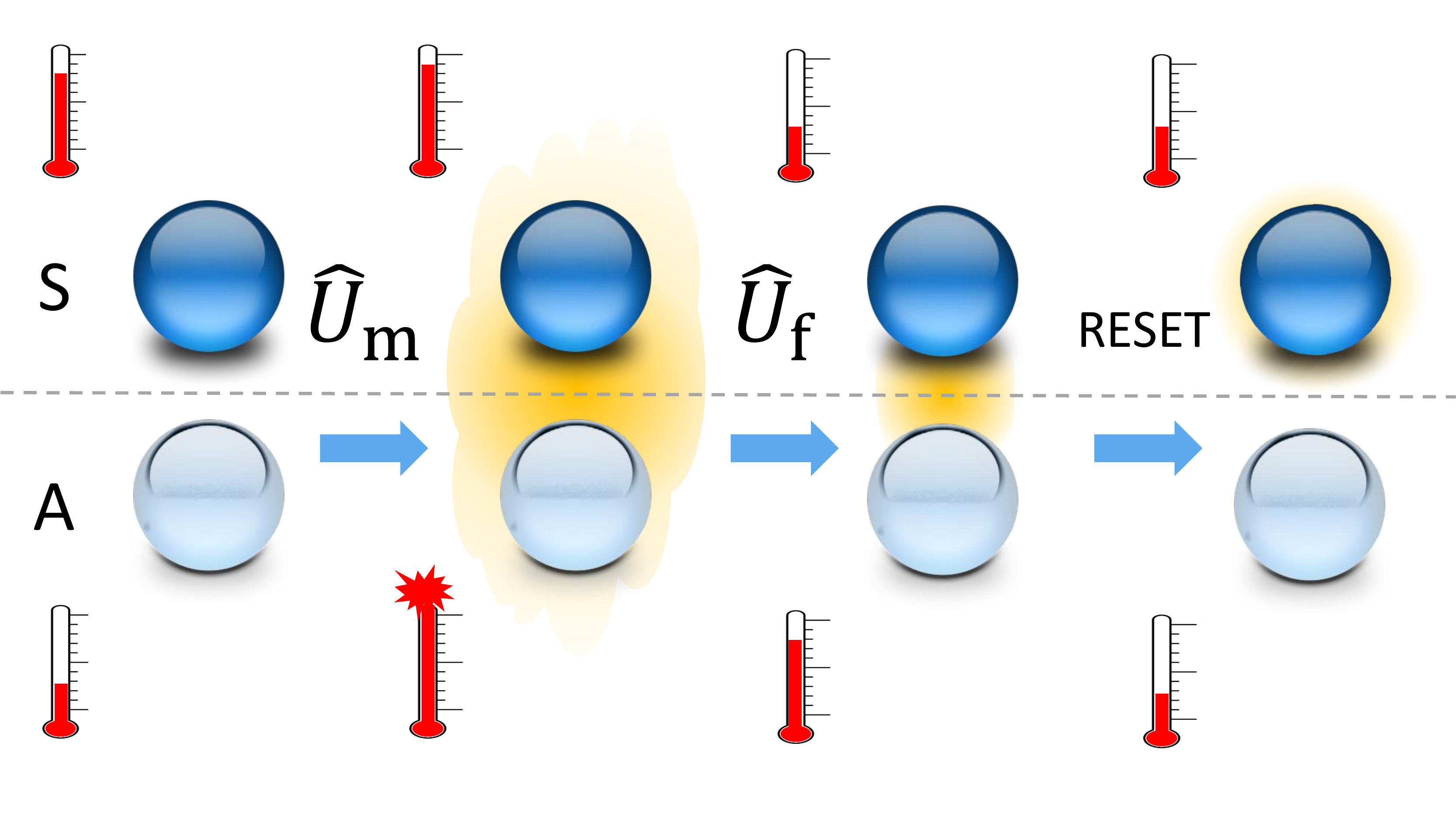}
	\caption{Sketch of the four steps of the spin cooling algorithm. The polarization bias of the marginals is illustrated by means of their effective ``spin temperatures'', indicated with thermometers, and the correlations and residual coherence are depicted as shaded yellow areas. First, $ \mathsf{S} $ and $ \mathsf{A} $ are initialized in an uncorrelated state with polarization biases $ \epsilon_\mathsf{S} < \epsilon_\mathsf{A} $. The measurement unitary $ \hat{U}_\text{m} $ correlates the two parts, yielding marginals with biases $ \epsilon_\mathsf{S}\cos{\varphi}^2 $ and zero, respectively (see the text for details on notation). After the application of the feedback unitary $ \hat{U}_\text{f} $, most correlations are wiped out as $ \mathsf{S} $ is mapped to the more polarized target $ \hat{\rho}_\mathsf{S} $, with polarization bias $ \epsilon_\mathsf{A}\sin{\varphi} $. The marginal of $ \mathsf{A} $ is then dissipatively reset to $ \hat{\rho}_0^{(\mathsf{A})} $.}
	\label{fig:diagram}
\end{figure}

During the protocol, the initial state $ \hat{\varrho}_0 = \hat{\rho}_0^{(\mathsf{S})}\otimes\hat{\rho}_0^{(\mathsf{A})} $ is mapped onto $ \hat{\varrho}_\text{f} $ under the application of $ \hat{U} $ (\textit{i.e.},~$\hat\varrho_0 \mapsto \hat{\varrho}_\text{f}\equiv\hat U \hat\varrho_0 \hat U^\dagger $). In light of coherent control, it seems insightful to split $ \hat{U} $ into two subsequent unitaries $\hat U_\text{m}$ and $\hat U_\text{f}$, so an intermediate ``post-measurement'' state $ \hat{\varrho}_\text{m} $ may be distinguished ($ \hat\varrho_0\mapsto \hat\varrho_\text{m} \equiv \hat U_\text{m}\hat\varrho_0 \hat U_\text{m}^\dagger\mapsto \hat \varrho_\text{f} \equiv \hat U_\text{f}\hat \varrho_\text{m} \hat U_\text{f}^\dagger $) \cite{PhysRevE.89.042134}. In the remainder of this section, we give further details on each of these steps (see also Figure~\ref{fig:diagram}).

\subsubsection{Initialization}\label{sec:protocol_initialization}

Initially, the two-level registers and ancillas are uncorrelated and have polarization biases \linebreak $0 < \epsilon_{\mathsf{S}} < \epsilon_{\mathsf{A}} < 1$, so that their joint state is $ \hat\varrho_0 = \hat\rho_0^{(\mathsf{S})}\otimes\hat\rho_0^{(\mathsf{A})} $, with marginals $ \hat\rho^{(\alpha)}_0=\frac12(\mathbb{I}_\alpha-\epsilon_\alpha\hat{\sigma}_z^{(\alpha)}) $ for $ \alpha\in\{\mathsf{S},\mathsf{A}\} $({in our notation, we indicate by the standard $\hat{\rho}^{(\alpha)}$ the marginal state of each subsystem $\alpha$ and by the variant $\hat{\varrho}$ the joint state of the whole system (registers plus ancillas), at any stage of the protocol}).
 Here, $ \hat{\sigma}_z^{(\alpha)} $ stands for the $ z $ Pauli matrix of subsystem $\alpha$. As already mentioned, we will assume that register and ancilla spins are all in thermal equilibrium with their surroundings at temperature $ T $, owing their difference in polarization to distinct energy gaps $ \hbar \omega_{\alpha} = k_B T \log{\big(\frac{1+\epsilon_\alpha}{1-\epsilon_\alpha}\big)} $. The global Hamiltonian may thus be written as $ \hat{H} = \hat{H}_\mathsf{S} + \hat H_\mathsf{A} = \frac{1}{2}{\hslash}\omega_\mathsf{S}\,\hat\sigma_z^{(\mathsf{S})}\otimes\mathbb{I}_\mathsf{A} + \mathbb{I}_\mathsf{S}\otimes\frac{1}{2}{\hslash}\omega_\mathsf{A}\,\hat{\sigma}_z^{(\mathsf{A})}$.

At this point, the von-Neumann entropies $ S({\hat{\rho}})\equiv -\text{tr}\{{\hat{\rho}}\log{{\hat{\rho}}}\} $ of the marginals \linebreak $ S({\hat{\rho}}_0^{(\mathsf{S})}) > S({\hat{\rho}}_0^{(\mathsf{A})})$ evaluate to:
\begin{equation}
S({\hat{\rho}}_0^{(\alpha)})=\frac12\log\left(\frac{4}{1-\epsilon_\alpha^2}\right)-\epsilon_\alpha\arctanh{\epsilon_\alpha}.
\label{eq:entropy_initial_marginal}
\end{equation}

\subsubsection{(Pre-)Measurement}

We wish to acquire information about the state of the registers by means of a quantum measurement. To that end, we can implement a joint unitary on $ \mathsf{S} + \mathsf{A} $, thus correlating the two~parties. The measurement unitary must be such that some information, about the local state of the registers (in~our case, its populations in some basis), gets imprinted on the marginal of the ancillas. In particular, our choice of measurement unitary $ \hat{U}_\text{m} $ will be \cite{PhysRevE.89.042134}:
\begin{equation}
\hat U_\text{m} = \exp{\left\{-i \frac{\pi}{4} \hat{\sigma}_{\vec{m}}^{(\mathsf{S})}\otimes\hat{\sigma}_y^{(\mathsf{A})}\right\}},
\label{eq:measurement_unitary}
\end{equation}
with $\hat\sigma_{\vec{m}}^{(\mathsf{S})}=\vec{m}\cdot\hat{\boldsymbol{\sigma}}^{(\mathsf{S})}$, $ \hat{\boldsymbol{\sigma}}^{(\mathsf{S})}=\{\hat\sigma_x^{(\mathsf{S})},\hat\sigma_y^{(\mathsf{S})},\hat\sigma_z^{(\mathsf{S})}\} $ and $|\vec{m}|=1$. This returns a state with marginals $ \hat\rho_\text{m}^{(\mathsf{S})}=\sum_{\mu{=\pm}}c_{\mu\mu}|\mu_{\vec{m}}\rangle\langle\mu_{\vec{m}}| $ and $ \hat\rho_\text{m}^{(\mathsf{A})}=\frac{1}{2}\mathbb{I}_\mathsf{A}-\frac{1}{2}\epsilon_{\mathsf{A}}\left(c_{++}-c_{--}\right)\hat\sigma_x^{(\mathsf{A})} $, where $|+_{\vec{m}}\rangle $ and $|-_{\vec{m}}\rangle$ are the eigenstates of $\hat\sigma_{\vec{m}}^{(\mathsf{S})}$ with eigenvalues $ +1 $ and $ -1 $, respectively, and $c_{\mu\mu}\equiv\left\langle\mu_{\vec{m}}\right\vert \hat{\rho}_0^{(\mathsf{S})}\left\vert\mu_{\vec{m}}\right\rangle$. That is, the coherences of the register in the eigenbasis of $ \hat{\sigma}_{\vec{m}}^{(\mathsf{S})} $ are destroyed, while the corresponding populations are recorded in the $ x $-basis of $ \mathsf{A} $ with `efficiency' $ \epsilon_\mathcal{\mathsf{A}} $. Hence, we can say that $ \hat{U}_\text{m} $ realizes an inefficient measurement of $ \hat{\sigma}_{\vec{m}}^{(\mathsf{S})} $ on $ \hat{\rho}_0^{(\mathsf{S})} $. Note also that given the initial cylindrical symmetry of the problem, we can restrict $ \vec{m} $ to the $ x $--
$ z $ plane, \textit{i.e.},~$ \vec{m}=\{\sin{\varphi},0,\cos{\varphi}\} $, with $ 0\leq\varphi\leq\frac{\pi}{2} $.

\subsubsection{Feedback}\label{sec:protocol_feedback}

Given the nature of $ \hat{U}_\text{m} $, the most informative measurement on $ \mathsf{A} $ about the state of $ \mathsf{S} $ is a projection onto the eigenstates $ \left\vert +_{\vec{x}} \right\rangle $ and $ \left\vert -_{\vec{x}} \right\rangle $ of $ \hat{\sigma}_x^{(\mathsf{A})} $. Therefore, conditioning the actuations of the controller on these measurement results has the potential to achieve the largest reduction in the entropy of the state of the register spins \cite{PhysRevE.89.042134}. In particular, our $ \hat{U}_\text{f} $ will take the form:
\begin{equation}
\hat{U}_\text{f}=\exp{\left\{i\frac{\pi}{4}\hat{\sigma}_y^{(\mathsf{S})}\right\}}\otimes\left\vert +_{\vec{x}} \right\rangle\left\langle +_{\vec{x}} \right\vert + \exp{\left\{-i\frac{\pi}{4}\hat{\sigma}_y^{(\mathsf{S})}\right\}}\otimes\left\vert -_{\vec{x}} \right\rangle\left\langle -_{\vec{x}} \right\vert.
\label{eq:feedback_unitary}
\end{equation}

After applying this feedback and regardless of $ \vec{m} $, the initial purity of the ancillas is transferred to the register spins (\textit{i.e}.,~$ \text{tr}\{{\rho_0^{(\mathsf{A})}}^{2}\} = \frac12(1+\epsilon_\mathsf{A}^2) = \text{tr}\{{\rho_\text{f}^{(\mathsf{S})}}^{2}\} $), while the entropy of the latter is reduced~by:
\begin{equation}
\Delta S^{(\mathsf{S})}_{0,\text{f}} \equiv S(\rho^{(\mathsf{S})}_0)- S(\rho^{(\mathsf{S})}_\text{f}) = \epsilon_\mathsf{A}\arctanh\epsilon_\mathsf{A}-\epsilon_\mathsf{S}\arctanh\epsilon_\mathsf{S} + \frac12\log{\left(\frac{1-\epsilon_\mathsf{A}^2}{1-\epsilon_\mathsf{S}^2}\right)} \geq 0.
\label{eq:entropy_reduction_feedback_marginal}
\end{equation}

In the extreme case of a measurement in the $ x $ direction (\textit{i.e}.,~$ \varphi = \pi/2 $ and $ \vec{m}=\{1,0,0\} $), the marginals of registers and ancillas are completely swapped after the application of $ \hat{U}_\text{f} $.

\subsubsection{Reset of the Ancilla}\label{sec:protocol_reset}

After a few relaxation times (measured in the time-scale of $ \mathsf{A} $), the irreversible interactions with the environment realize the transformation $ \hat\varrho_\text{f} \mapsto \hat\rho_\text{f}^{(\mathsf{S})}\otimes\hat\rho_0^{(\mathsf{A})} $, destroying all of the correlations between the register and the controller. The system will be then ready in principle for additional rounds of feedback cooling, always provided that the polarization bias of the ancillary spins can be still increased. In the specific implementation considered in this paper, however, given our choice of $ \hat{U}_\text{m} $ and $ \hat{U}_\text{f} $, any further increase in the bias of the registers or reduction of their entropy will be in fact impossible. For this reason, in all that follows, we shall consider only a single iteration of the~protocol.


\section{Thermodynamic Analysis}\label{sec:thermodynamics}

\subsection{Energy Balance}\label{sec:thermodynamics_energy}

Let us start by assessing the energetics of the measurement step. The application of the unitary $ \hat{U}_\text{m} $ has always a negative energy cost for the controller, \textit{i.e}.,~net {\it work} must be performed,
\begin{equation}
\Delta E_{0,\text{m}}\equiv\text{tr}\{\hat{H}(\hat{\varrho}_0-\hat{\varrho}_\text{m})\} = - k_B T(\epsilon_\mathsf{S}\sin^2{\varphi}\arctanh{\epsilon_\mathsf{S}}+\epsilon_\mathsf{A}\arctanh{\epsilon_\mathsf{A}}) < 0.
\label{eq:work_measurement}
\end{equation}

Interestingly, during the feedback, it may be possible for the controller to recover a fraction of the work invested in the measurement. Actually, for choices of $ \vec{m} $ with $ \varphi $ above a certain threshold $ \varphi_\text{crit} $, the feedback unitary of Equation \eqref{eq:feedback_unitary} always succeeds at extracting work from $ \hat\varrho_\text{m} $, besides minimizing the marginal entropy of $ \mathsf{S} $. {In fact, we have work extraction if:
\begin{equation}
 \Delta E_{\text{m},\text{f}}\equiv\text{tr}\{\hat{H}(\hat{\varrho}_\text{m}-\hat{\varrho}_\text{f})\} =- k_B T \left(y\sin{\varphi}-\epsilon_\mathsf{S}\arctanh{\epsilon_\mathsf{S}}\cos^2{\varphi}\right) \geq0
\end{equation}
where $ y \equiv \epsilon_\mathsf{A}\arctanh{\epsilon_\mathsf{S}}+\epsilon_\mathsf{S}\arctanh{\epsilon_\mathsf{A}} $. The threshold value $ \varphi_\text{crit} $ is thus given by}:
\begin{equation}
\sin{\varphi_\text{crit}}=\frac{-y+\sqrt{y^2+4\epsilon_\mathsf{S}^2\arctanh^2{\epsilon_\mathsf{S}}}}{2\epsilon_\mathsf{S}\arctanh{\epsilon_\mathsf{S}}}
\label{eq:varphi_work_extr}
\end{equation}

{In particular, the maximum work recovery $ \Delta E_{\text{m},\text{f}} $} is attained for measurements along the $ x $ direction $ (\varphi=\pi/2) $, although not all extractable work (or ``ergotropy'' \cite{allahverdyan2004maximal}) can be retrieved.

Finally, during the reset, the ancillary spins lose an amount of {\it heat}:
\begin{equation}
 \mathcal{Q} \equiv \text{tr}\{\hat H_\mathsf{A}(\hat{\rho}_\text{f}^{(\mathsf{A})}-\hat{\rho}^{(\mathsf{A})}_0) \}=k_B T(\epsilon_\mathsf{A}-\epsilon_\mathsf{S}\sin{\varphi})\arctanh{\epsilon_\mathsf{A}} > 0
\label{eq:heat_reset}
\end{equation}
irreversibly to the environment, while the residual correlations between $ \mathsf{S} $ and $ \mathsf{A} $ are completely erased, and the marginal state of the registers remains unchanged. Hence, the ancillary spins perform a cycle, whereas the registers change their average energy by:
\begin{equation}
\Delta E_{0,\text{f}}^{(\mathsf{S})}=\text{tr}\{ \hat{H}_\mathsf{S}(\hat{\rho}_0^{(\mathsf{S})}-\hat{\rho}_\text{f}^{(\mathsf{S})}) \} = -k_B T(\epsilon_\mathsf{S}-\epsilon_\mathsf{A}\sin{\varphi})\arctanh{\epsilon_\mathsf{S}}
\label{eq:system_cooling}
\end{equation}

Note that $ \Delta E_{0,\text{f}}^{(\mathsf{S})} $ does not have a definite sign. Indeed, due to the residual coherence in $ \hat{\rho}_\text{f}^{(\mathsf{S})} $, the increase in the system's purity achieved in the protocol does not have to be accompanied by an increase in the polarization bias of the register spins, or equivalently, by a decrease in their average energy. This would additionally require that $ \epsilon_\mathsf{A}\sin{\varphi} > \epsilon_\mathsf{S}$, which is only guaranteed to hold for a measurement along the eigenbasis of $ \hat{\sigma}_x^{(\mathsf{S})} $. Put in other words, even though $ \Delta S_{0,\text{f}}^{(\mathsf{S})} $ is always non-negative by construction of the algorithm, \textit{real cooling} of the registers only happens within the `cooling window' $ \frac{\epsilon_\mathsf{S}}{\sin{\varphi}} < \epsilon_\mathsf{A} \leq 1 $. An important consequence of this is that the polarization bias of the registers cannot be increased if $ \sin{\varphi} < \epsilon_\mathsf{S} $, regardless of $ \epsilon_\mathsf{A} $.

\subsection{Performance of Feedback Cooling}\label{sec:thermodynamics_performance}

Performance optimization is a key element in the study of thermodynamic energy-conversion cycles. It is essential to find that {sweet spot} in the parameter space of the cycle in question, signalling the most energy-efficient usage of the input resources conditioned on the maximization of the ``useful effect''. Depending on the situation, one may be willing to spend more resources in order to achieve, e.g.,~faster cooling, or to minimize any undesired side-effects, such as residual heating of the environment. In each case, the relevant trade-off can be captured by a suitably-defined figure of~merit.

In particular, our feedback cooling algorithm was designed to minimize the entropy of the register spins so that one may identify $\mathcal{P} \equiv k_B T \Delta S_{0,\text{f}}^{(\mathsf{S})} $ as the useful effect (see Equation \eqref{eq:entropy_reduction_feedback_marginal}). \linebreak On the other hand, the total work penalty for the controller is given by $ \mathcal{W} \equiv - \text{tr}\{\hat{H}(\hat{\varrho}_0-\hat{\varrho}_\text{f})\} = -\Delta E_{0,\text{f}}^{(\mathsf{S})} + \mathcal{Q} > 0 $ (\textit{cf}. Equations \eqref{eq:system_cooling} and \eqref{eq:heat_reset}), so that we can define the \textit{coefficient of performance} (COP) of the protocol as the quotient $ \varepsilon\equiv\mathcal{P}/\mathcal{W} $. This COP is positive and unbounded, just like the ratio of the cooling rate to the input power in a conventional refrigeration cycle \cite{Gordon2000}.

\begin{figure}[H]
	\centering
	\includegraphics[scale=0.5] 	{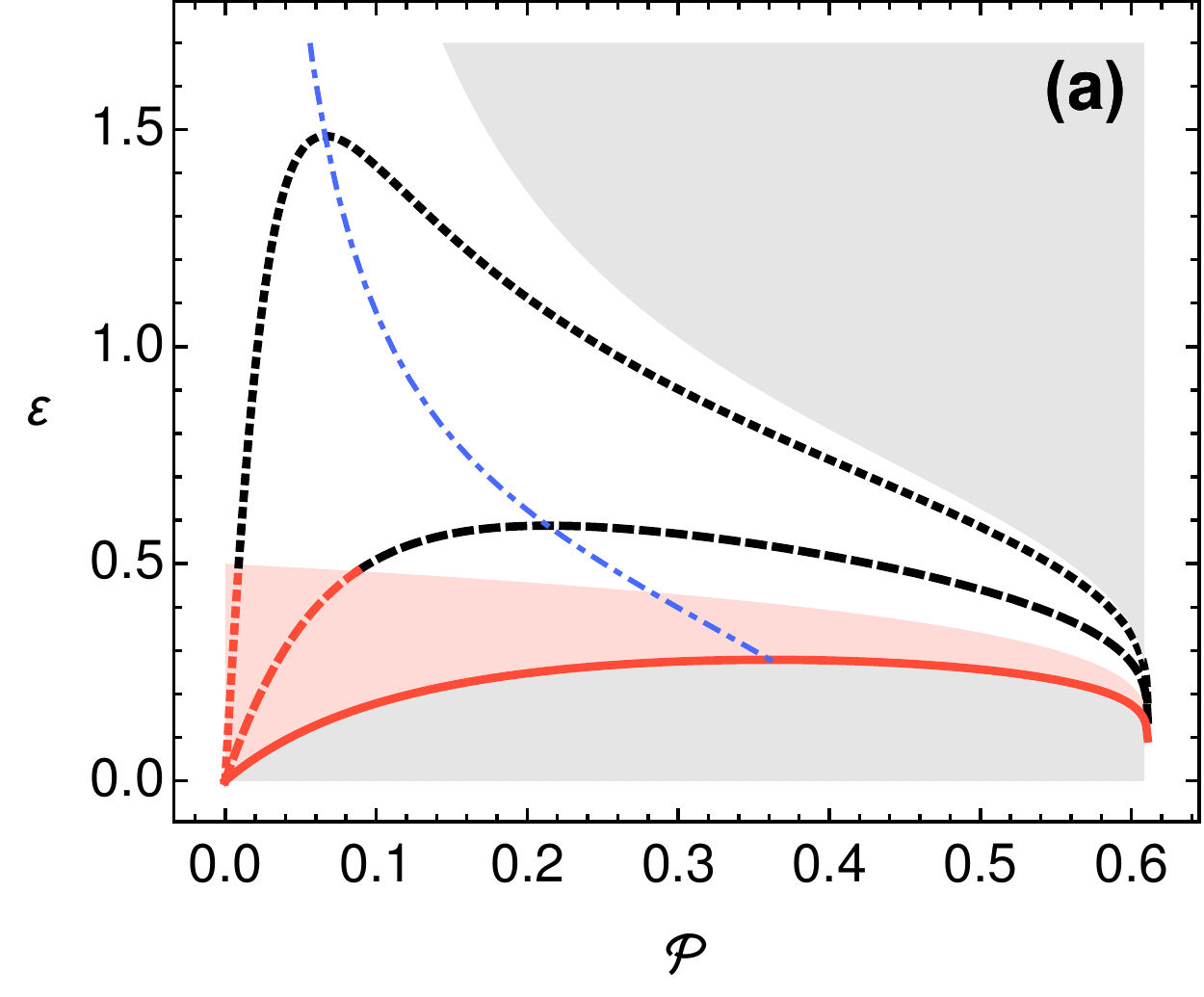}
\hspace*{.5cm}		\includegraphics[scale=0.5]	{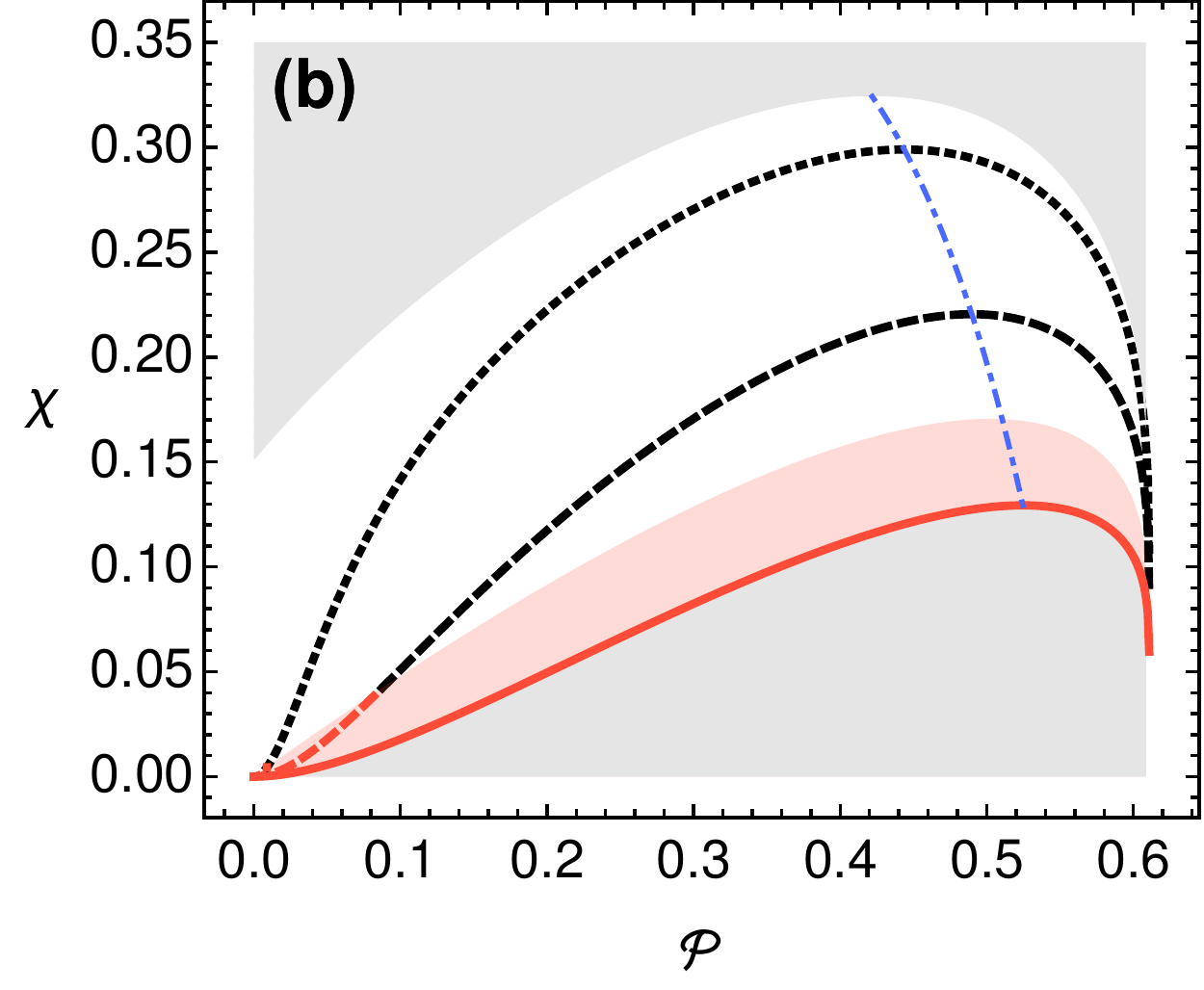}
	\caption{(\textbf{a}) Coefficient of performance and (\textbf{b}) figure of merit $ \chi $ \textit{versus} the entropy reduction on the registers $ \mathcal{P} = k_B T \Delta S_{0,\text{f}}^{(\mathsf{S})} $ for fixed initial polarization bias $ \epsilon_\mathsf{S} = 0.4 $ and different measurement directions: $ \varphi = 0 $ (solid), $ \varphi = \pi/4 $ (dashed) and $ \varphi = 2\pi/5 $ (dotted). In both plots, the bias of the ancillas $ \epsilon_\mathsf{A} $ ranges from $ \epsilon_\mathsf{S} $ to one, and the temperature is $ T = 1 $. The part of the curves falling inside the cooling window $ \frac{\epsilon_\mathsf{S}}{\sin{\varphi}} < \epsilon_\mathsf{A} < 1 $ is depicted in black, whereas configurations for which $ \Delta E_{0,\text{f}}^{(\mathsf{S})} < 0 $ (\textit{i.e.},~no real cooling occurs) lie within the shaded red areas. The grey regions correspond to inaccessible configurations, and the optimal working points $ \{\mathcal{P}^\star,\varepsilon^\star\} $ and $ \{\mathcal{P}^\star,\chi^\star\} $ are indicated with dot-dashed blue lines.}
\label{fig:COP}
\end{figure}

A plot of $ \varepsilon $ \textit{versus} $ \mathcal{P} $ for fixed $ \epsilon_\mathsf{S} $ and varying $ \epsilon_\mathsf{A} $ can be thought of as the ``characteristic curve''~\cite{gordon1991generalized} of our feedback cooling cycle, which is sketched in Figure \ref{fig:COP}a. Given a measurement direction $ \vec{m} $, the COP is maximized at an intermediate polarization bias $ \epsilon_\mathsf{S} < \epsilon_{\mathsf{A}}^\star < 1 $ corresponding to some optimal entropy reduction rate $ \mathcal{P}^\star $. The value of $ \mathcal{P}^\star $ decreases as $ \vec{m} $ sweeps from $ \{0,0,1\} $ to $ \{1,0,0\} $, while the COP grows monotonically for any fixed $ \epsilon_\mathsf{A} $. In the limiting case of an $ x $-measurement ($ \varphi = \pi/2 $), $ \varepsilon $ attains its maximum value as $ \epsilon_\mathsf{A} \rightarrow \epsilon_\mathsf{S} $, although at vanishing $ \mathcal{P}^\star $. \linebreak Recall from Equation \eqref{eq:heat_reset} that $ \varphi = \pi/2 $ and $ \epsilon_\mathsf{A} \rightarrow \epsilon_\mathsf{S} $ entail $ \mathcal{Q} = 0 $, which means that the maximization of the COP occurs when the overall operation is reversible, as should be expected.

In~\cite{PhysRevE.89.042134}, an alternative efficiency-like objective function $ \eta \equiv \mathcal{P}/\mathcal{Q} $ had been introduced. Its optimization guarantees the largest entropy reduction at the expense of minimal heat release into the environment. From Equations \eqref{eq:entropy_reduction_feedback_marginal} and \eqref{eq:heat_reset}, it follows that $ \mathcal{Q} \geq \mathcal{P} $. As a result, $ \eta $ is positive and upper-bounded by one. Not surprisingly, its qualitative behaviour is the same as that of the COP. \linebreak Note that the regime of operation in which the feedback unitary becomes capable of extracting work from $ \hat{\varrho}_\text{m} $ (\textit{cf}. Section~\ref{sec:thermodynamics_energy}) is very relevant from the point of view of performance optimization: not only the net work supplied by the controller is reduced, but also the excess energy dissipated as heat during the reset. We shall come back to this point in Section~\ref{sec:information}.

From all of the above, it is clear that the thermodynamically optimal feedback cooling protocol must start with the measurement of the polarization bias of the register spins in the eigenbasis of $ \hat{\sigma}_x^{(\mathsf{S})} $. \linebreak As already mentioned, in that case, the overall unitary manipulation $ \hat{U} $ simply amounts to swapping the states of $ \mathsf{S} $ and $ \mathsf{A} $. However, the fact that the global maximum of both the COP and the efficiency $ \eta $ is met as $ \epsilon_\mathsf{A}\rightarrow\epsilon_\mathsf{S} $ and at a vanishing entropy reduction rate may seem unsatisfactory. Even though this is very often the case in thermal engineering ({this happens, for instance, in an} \textit{endoreversible} \cite{hoffmann1997endoreversible} {refrigerator model with linear heat transfer laws }\cite{0022-3727_23_2_002}, {which is a classical case study in finite-time thermodynamics}),
 one would wish instead for a meaningful figure of merit attaining its maximum at a non-vanishing (and ideally large) ``cooling load''.

To resolve this issue, we can introduce the objective function $ \chi\equiv\varepsilon\mathcal{P} $, which is well suited for applications in which maximizing the cooling load is as important as increasing the energy efficiency of the cycle \cite{0022-3727_23_2_002,PhysRevE.85.010104,PhysRevE.86.011127}. As we can see in Figure~\ref{fig:COP}b, $ \chi $ is qualitatively different from $ \varepsilon $ and $ \eta $: Its global maximum is still attained for $ \varphi = \pi/2 $, but the corresponding entropy reduction rate $ \Delta S_{0,\text{f}}^{(\mathsf{S})\,\star} $ remains comparatively large, which is of practical importance. Using the figure of merit $ \chi $ and Equations~\eqref{eq:entropy_reduction_feedback_marginal}--\eqref{eq:heat_reset}, it is easy to see that the best compromise between entropy reduction and work expenditure is attained when the ancillary spins are prepared at a large polarization bias ($ \epsilon_\mathsf{A}^\star \gtrapprox 0.8 $).


\section{Information-Theoretic Analysis}\label{sec:information}

Up to now, the breakdown of $ \hat{U} $ into two distinct steps $ \hat{U}_\text{m} $ and $ \hat{U}_\text{f} $ does not seem to have added anything relevant to the discussion, since the balance of the input/output energy and the entropy reduction in $ \mathsf{S} $ have been both evaluated globally. However, adopting the viewpoint of measurement-based quantum feedback control can shed new light on the problem if the \textit{information balance} is considered instead. Recall that the measurement unitary is designed so that the controller acquires information about the system by correlating the registers with the ancillary spins in a suitable way. The subsequent actions of the controller on the register spins, and the ultimate success of the feedback cooling cycle are conditioned on that information. It thus seems interesting to gauge the build-up of correlations during the measurement step with quantitative measures and, in particular, to tell apart their quantum share from the classical one. This study, side by side with our assessment of the energetics of the protocol, will help to establish connections between genuinely non-classical effects and the thermodynamic optimization of quantum feedback cooling algorithms.

We already know from Section~\ref{sec:thermodynamics_performance} that the energy efficiency is maximized when the measurements are carried out in the eigenbasis of $ \hat{\sigma}_x^{(\mathsf{S})} $. This is a basis with respect to which the initial thermal state $ \hat{\rho}_0^{(\mathsf{S})} $ has maximal coherence, and therefore, performing an $ x $-measurement can be regarded as the most ``quantum'' instance of the cooling protocol. In contrast, measuring the system in the energy eigenbasis amounts to the only completely ``classical'' situation. The fact that the latter realizes the worst-case scenario when it comes to performance seemingly indicates that some element of quantumness may be a resource for the algorithm, as suggested in~\cite{PhysRevE.89.042134}. \linebreak Interestingly, quantum correlations are also known to increase the extractable work in control loops with quantum feedback \cite{park2013heat}.

In what follows, we will try to make this intuition precise by computing the quantum correlations in the form of \textit{entanglement} \cite{RevModPhys.81.865} and \textit{quantum discord} \cite{olliver20011,henderson20011} between $ \mathsf{S} $ and $ \mathsf{A} $, immediately after the measurement step.

\subsection{Entanglement}\label{sec:information_entanglement}

Quantum entanglement is the most popular signature of non-classicality in bipartite quantum states. Simply put, the state $ \hat{\varrho} $ is entangled if it cannot be written as $\hat{\varrho}\neq\sum_i p_i\hat{\rho}_i^{(\mathsf{S})}\otimes\hat{\rho}_i^{(\mathsf{A})}$, where $ p_i $ is some probability distribution and $ \hat{\rho}_i^{(\alpha)} $ are local states of $ \mathsf{S} $ and $ \mathsf{A} $. In other words, entangled states cannot be prepared by means of local operations and classical communication between the two~parties.

Entanglement is rooted at the very heart of quantum theory \cite{schrodinger1935gegenwartige} and has been paramount in historical controversies, such as the Einstein--Podolsky--Rosen argument \cite{einstein1935can} on the alleged incompleteness of quantum theory and the latter resolution of the issue by Bell, showing the incompatibility of quantum theory and any local hidden-variable model \cite{bell1964einstein}. However, the main reason for the popularity of entanglement over the last couple of decades is its role as a resource for quantum technologies, enabling, e.g.,~quantum teleportation \cite{PhysRevLett.70.1895,bouwmeester1997experimental}, better-than-classical information processing \cite{bennett19921,PhysRevLett.76.4656}, quantum cryptography \cite{ekert19911,PhysRevLett.84.4729} or enhanced quantum metrology \cite{PhysRevLett.79.3865}.

Amongst all available quantifiers of entanglement \cite{RevModPhys.81.865}, we shall use the \textit{entanglement of formation}, which, for two-qubit states, reads \cite{PhysRevLett.80.2245}:
\begin{equation}
\mathcal{E}(\hat{\varrho})\equiv h\left(\frac{1+\sqrt{1-C(\hat{\varrho})^2}}{2}\right),
\label{eq:entanglement_of_form}
\end{equation}
with $ h(x) \equiv -x\log{x}-(1-x)\log{(1-x)} $. The concurrence $ C(\hat{\rho}) $ can be computed from:
\begin{equation}
C(\hat{\varrho})=\max{\{0,2\lambda_{\text{max}}-\text{tr}\,\hat{R}(\hat{\varrho})\}},
\label{eq:concurrence}
\end{equation}
where $\lambda_{\text{max}}$ is the largest eigenvalue of the operator $\hat{R}(\hat{\varrho})$, defined as:
\begin{equation}
\hat{R}(\hat{\varrho})=\sqrt{\sqrt{\hat{\varrho}}\,(\hat{\sigma}_y^{(\mathsf{S})}\otimes\hat{\sigma}_y^{(\mathsf{A})})\,\hat{\varrho}^*\,(\hat{\sigma}_y^{(\mathsf{S})}\otimes\hat{\sigma}_y^{(\mathsf{A})})\sqrt{\hat{\varrho}}}.
\label{eq:concurrence_aux}
\end{equation}

\begin{figure}[H]
	\centering
	\hspace*{-.2cm}\includegraphics[scale=0.49] %
	{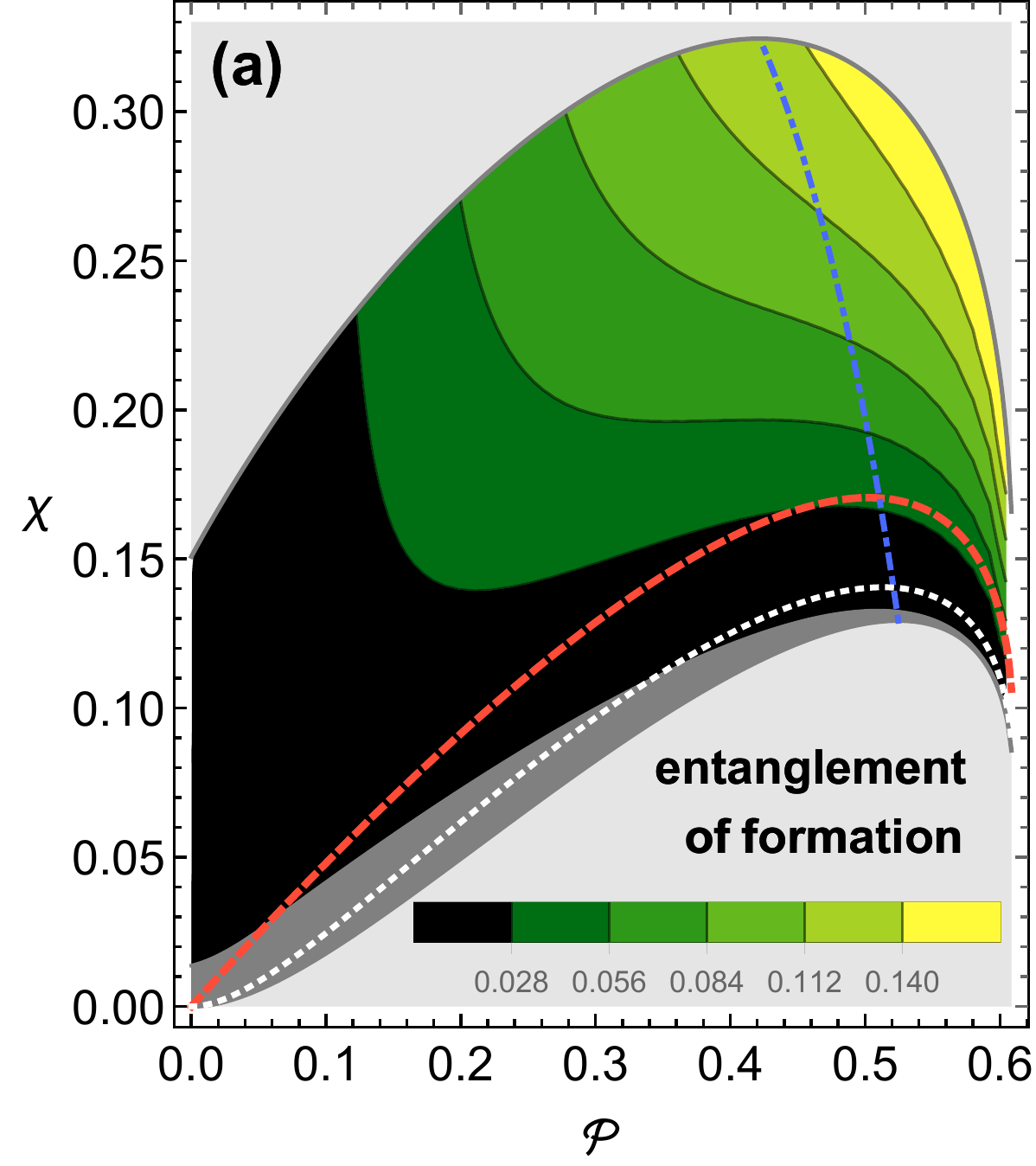}
	\includegraphics[scale=0.49]%
	{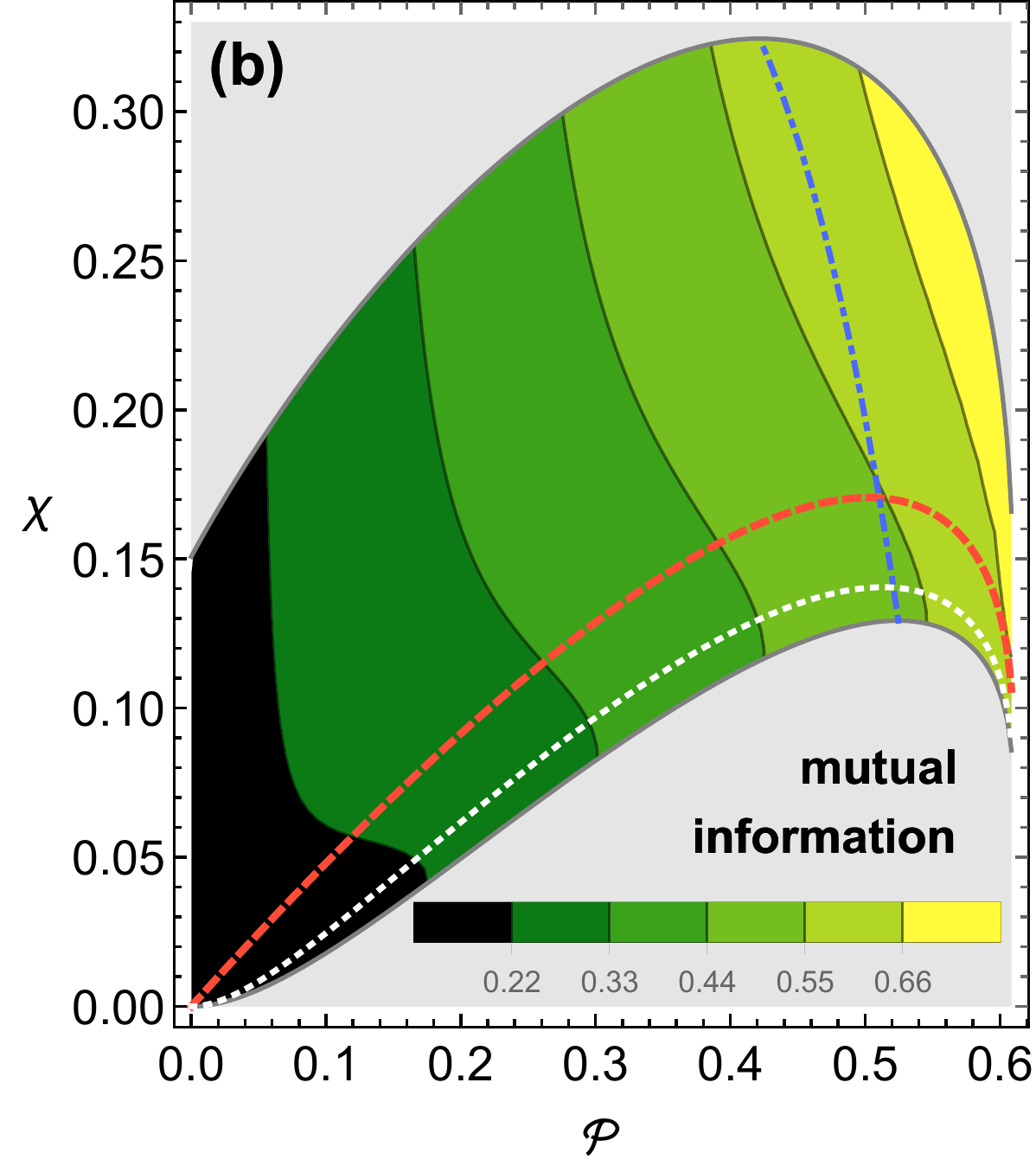}\\
	\includegraphics[scale=0.49]%
	{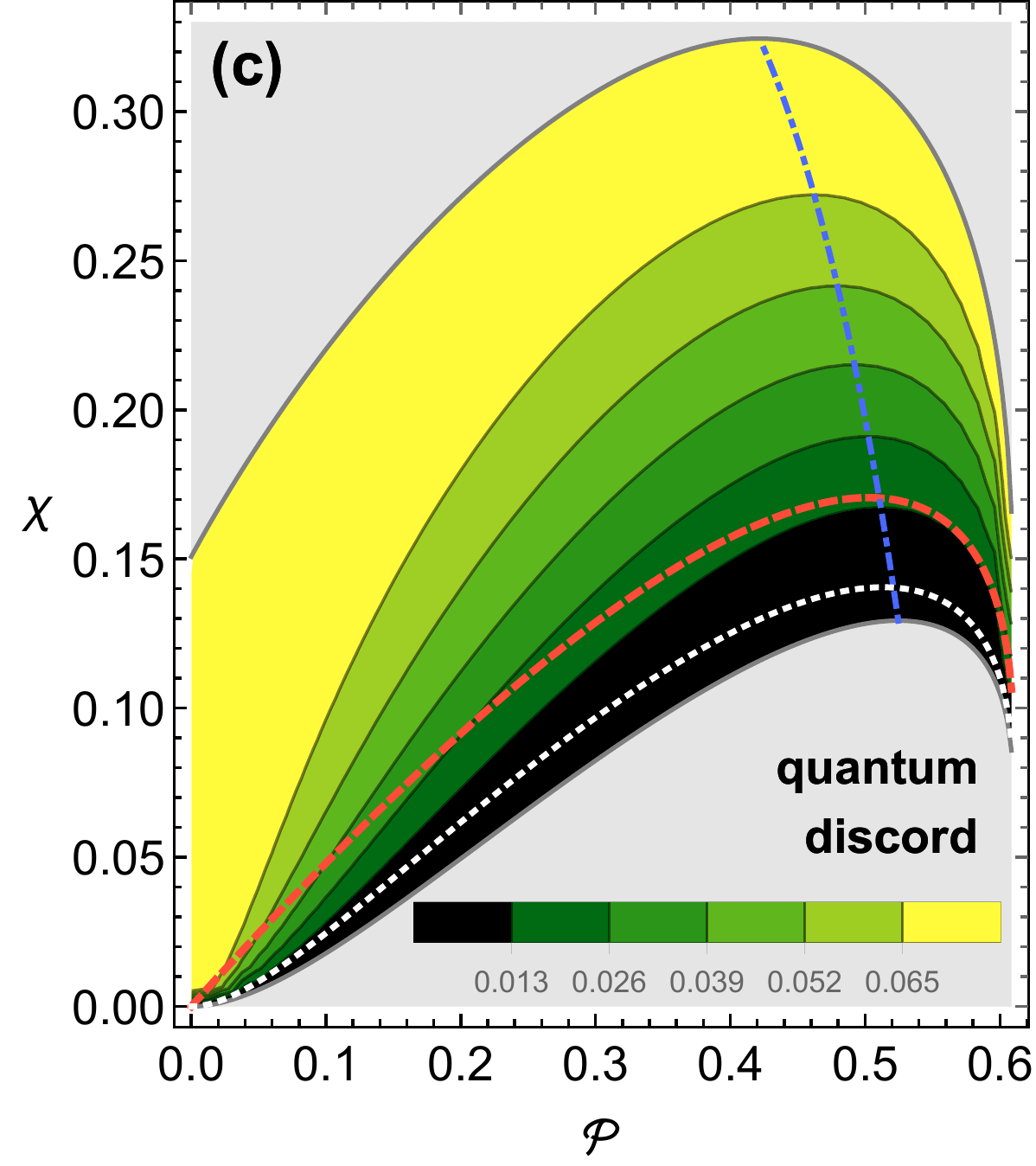}
	\caption{(\textbf{a}) Entanglement of formation $ \mathcal{E}(\hat{\varrho}_\text{m}) $, (\textbf{b}) mutual information $ I(\hat{\varrho}_\text{m}) $ and (\textbf{c}) quantum discord $ \delta_\mathsf{A}(\hat{\varrho}_\text{m}) $ evaluated after the measurement step, \textit{versus} the entropy reduction rate $ \mathcal{P} $ and the figure of merit $ \chi $. As in Figure~\ref{fig:COP}, the shaded grey areas, the dashed red curve and dot-dashed blue curve correspond to inaccessible configurations, the threshold towards effective cooling and the optimal operation points, respectively. The dotted white line marks configurations {above} which the feedback unitary $ \hat{U}_\text{f} $ becomes capable of extracting work from $ \hat{\rho}_\text{m} $ (\textit{cf.} Equation~\eqref{eq:varphi_work_extr}). Finally, the dark shaded grey area of {(a)} corresponds to working points with zero entanglement between $ \mathsf{S} $ and $ \mathsf{A} $. We have set $ \epsilon_\mathsf{S} = 0.4 $ and $ T = 1 $.}
	\label{fig:correlations}
\end{figure}

In Figure~\ref{fig:correlations}a, we plot the entanglement of formation for all working points $ \{{\chi}, \mathcal{P}\} $. We see that setups with low enough $ \varphi $ produce separable post-measurement states $ \hat{\varrho}_\text{m} $ (falling within the dark shaded grey area), although entanglement is almost ubiquitous in this protocol.

We can also observe that entanglement may not be directly linked with the ability of the cycle to increase the polarization bias of the registers: depending on the entropy reduction rate considered, both separable and entangled states $ \hat{\varrho}_\text{m} $ may succeed or fail to increase the polarization bias of the registers after the feedback step. Exactly the same can be said about the potential relation between $ \mathcal{E}(\hat{\varrho}_\text{m}) $ and the possibility of work extraction ($ \Delta E_{\text{m},\text{f}} > 0 $) by $ \hat{U}_\text{f} $ from the post-measurement state: entanglement between $ \mathsf{S} $ and $ \mathsf{A} $ is definitely not a necessary ingredient \cite{WorkEnt}.

Furthermore, the overall maximization of $ \mathcal{E}(\hat{\varrho}_\text{m}) $ occurs as $ \{\varphi,\epsilon_\mathsf{A}\}\rightarrow\{\frac{\pi}{2},1\} $, which is far from being an optimal working point of the cycle. At most, one can say that, fixing $ \mathcal{P} $, the entanglement between registers and ancillas after the measurement is a monotonically increasing function of the figure of merit $ \chi $ (as well as of $ \varepsilon $ and $ \eta $). However, the same can be said about the total, classical and quantum correlations built up in $ \hat{\varrho}_\text{m} $, as we shall see next.

\subsection{Total, Quantum and Classical Correlations}\label{sec:information_discord}

To begin with, recall that any state $ \hat{\varrho}\neq\hat{\rho}^{(\mathsf{S})}\otimes\hat{\rho}^{(\mathsf{A})} $ is said to be correlated and that its total correlation content can be measured with the quantum mutual information:
\begin{equation}
I(\hat{\varrho})\equiv S(\hat{\rho}^{(\mathsf{S})}) + S(\hat{\rho}^{(\mathsf{A})}) - S(\hat{\varrho})
\label{eq:mutual_information}
\end{equation}

Now, let us consider a separable state of the form $ \hat{\varrho} = \sum_i p_i \, \hat{\rho}_i^{(\mathsf{S})}\otimes\hat{\Pi}_i^{(\mathsf{A})} $, where $ \hat{\Pi}_i^{(\mathsf{A})} \equiv \left\vert i_\mathsf{A} \right\rangle\left\langle i_\mathsf{A} \right\vert $ is a shorthand for the projectors onto the non-degenerate eigenstates of some observable $ \hat{O}_\mathsf{A} $ acting on $ \mathsf{A} $. In this particular case, one can alternatively write the mutual information as:
\begin{equation}
I(\hat{\varrho}) = S(\hat{\rho}^{(\mathsf{S})}) - S(\hat{\varrho}\vert\{\hat{\Pi}_i^{(\mathsf{A})}\}),
\label{eq:mutual_information_2}
\end{equation}
where $ S(\hat{\varrho}\vert\{\hat{\Pi}_i^{(\mathsf{A})}\}) \equiv \sum_i S\big( \hat{\Pi}_i^{(\mathsf{A})}\hat{\varrho}\hat{\Pi}_i^{(\mathsf{A})}\big) = \sum_i p_i\,S(\hat{\rho}_i^{(\mathsf{S})})$ is the average of the entropies of the states $ \hat{\Pi}_i^{(\mathsf{A})}\hat{\varrho}\hat{\Pi}_i^{(\mathsf{A})} / \text{tr}\{\hat{\Pi}_i^{(\mathsf{A})}\hat{\varrho}\}$, resulting from a local measurement of $ \hat{O}_\mathsf{A} $, weighted by the corresponding probabilities for each measurement outcome (\textit{i.e.},~$ \text{tr}\{\hat{\Pi}_i^{(\mathsf{A})}\hat{\varrho}\} $). However, the existence of a complete local projective measurement, which leaves the marginal $ \hat{\rho}_\text{m}^{(\mathsf{S})} $ unperturbed (that is, such that $ \text{tr}_\mathsf{A}\{\sum_i \hat{\Pi}_i^{(\mathsf{A})}\hat{\varrho}\hat{\Pi}_i^{(\mathsf{A})}\} = \hat{\rho}_\text{m}^{(\mathsf{S})} $), is a specific property of our state $ \hat{\varrho} = \sum_i p_i \, \hat{\rho}_i^{(\mathsf{S})}\otimes\hat{\Pi}_i^{(\mathsf{A})} $. Such measurement usually cannot be found for an arbitrary bipartite separable state. Consequently, one generally finds that $ I(\hat{\varrho}) > S(\hat{\rho}^{(\mathsf{S})}) - S(\hat{\varrho}\vert\{\hat{\Pi}_i^{(\mathsf{A})}\}) $ for any choice of $ \{\hat{\Pi}_i^{(\mathsf{A})}\} $.

One may say that the fact that {any} local measurement on $ \mathsf{A} $ produces a disturbance on the marginal of $ \mathsf{S} $ is a genuinely non-classical feature of the bipartite state $ \hat{\varrho} $. One may quantify this departure from classicality by defining the {\it quantum discord} \cite{olliver20011} $ \delta_\mathsf{A}(\hat{\varrho}) $ to be:
\begin{equation}
\delta_\mathsf{A}(\hat{\rho})\equiv I(\hat{\varrho})-\sup_{\{\hat{\Pi}_i^{(\mathsf{A})}\}}\left[ S(\hat{\rho}^{(\mathsf{S})}) - S(\hat{\varrho}\vert\{\hat{\Pi}_i^{(\mathsf{A})}\}) \right].
\label{eq:discord_definition}
\end{equation}

One may as well define the \textit{classical share} of correlations simply as $ \mathcal{C}_\mathsf{A}(\hat{\varrho}) = I(\hat{\varrho})-\delta_\mathsf{A}(\hat{\varrho}) $ \cite{henderson20011}. Note that both $ \delta $ and $ \mathcal{C} $ are generally asymmetric, e.g.,~even if the state $ \hat{\varrho} = \sum_i p_i \, \hat{\rho}_i^{(\mathsf{S})}\otimes\hat{\Pi}_i^{(\mathsf{A})} $ is classically correlated with respect to local measurements on $ \mathsf{A} $, it may have non-zero discord with respect to measurements on $ \mathsf{S} $ (\textit{i.e.},~$ \delta_S(\hat{\varrho}) > 0 $). Most importantly, note that even if all entangled states are discordant, non-zero discord may be readily found also in separable states. For this reason, discord is said to be an indicator of quantumness beyond entanglement.

Several alternative measures of discord have been proposed over the last few years \cite{RevModPhys.84.1655}, and the role of discord-type correlations in applications, such as quantum communication, cryptography and metrology, has been investigated \cite{operd1,operd2,operdc,PhysRevLett.110.240402,PhysRevLett.112.210401}. From a thermodynamic perspective, discord captures the additional work extractable from correlated quantum systems by quantum Maxwell's demons, as compared to classical ones \cite{operdz}.

In Figure~\ref{fig:correlations}b, we represent the quantum mutual information $ I(\hat{\varrho}_\text{m}) $. We shall omit here its lengthy analytical expression, which is reported in the Appendix. Instead, we will only note that, similarly to the entanglement, the total and classical correlations are globally maximized as $ \{\varphi,\epsilon_\mathsf{A}\}\rightarrow\{\frac{\pi}{2},1\} $, which does not coincide with the optimal working points according to neither $ \chi $ nor $ \varepsilon $ or $ \eta $.

Interestingly, the situation is quite different when the quantum share of correlations is considered instead, as illustrated in Figure~\ref{fig:correlations}c. While its evaluation is challenging in general, the quantum discord in our post-measurement state $ \hat{\varrho}_\text{m} $ can be evaluated analytically from the definition in Equation~\eqref{eq:discord_definition}:
\begin{equation}
\delta_{\mathsf{A},\mathsf{S}}(\hat{\varrho}_\text{m})\equiv\delta(\hat{\varrho}_\text{m})=\epsilon_\mathsf{S}\arctanh\epsilon_\mathsf{S}+\frac{1}{2}\log{\left(\frac{\epsilon_\mathsf{S}^2-1}{\epsilon_\mathsf{S}^2\cos^2{\varphi}-1}\right)}+\frac{1}{2}\epsilon_\mathsf{S}\cos{\varphi}\log{\left(\frac{1-\epsilon_\mathsf{S}\cos{\varphi}}{1+\epsilon_\mathsf{S}\cos{\varphi}}\right)}.
\label{eq:discord_m}
\end{equation}

As we can see, the quantum correlations between registers and ancillas after the measurement step are \textit{symmetric} under the exchange of the parties. It is also noteworthy that $ \delta(\hat{\varrho}_\text{m})$, at variance with the entropy reduction rate $ \mathcal{P} $, does not depend on the polarization bias of the ancillary spins $ \epsilon_\mathsf{A} $, but only on $ \epsilon_\mathsf{S} $ and the measurement direction $ \varphi $. Crucially, this implies that the performance characteristics introduced in Section~\ref{sec:thermodynamics_performance} are all \textit{curves of constant discord}. In particular, the maximization of $ \delta(\hat{\varrho}_\text{m}) $ at fixed $ \epsilon_\mathsf{S} $ occurs now as $ \varphi\rightarrow\frac{\pi}{2} $, which is compatible with the optimization of the objective function $ \chi $, as well as that of the COP and $ \eta $. This observation is suggestive of a deep connection between quantum discord and the thermodynamic performance of the spin refrigeration cycle: while all figures of merit increase monotonically with the total, classical and entangled correlations at fixed entropy production rate, only the quantum share of correlations is seen to grow monotonically with $ \chi $, $ \varepsilon $ and $ \eta $ at {any} cooling load $ \mathcal{P} $.

The red-dashed and white-dotted lines in Figure~\ref{fig:correlations}c that delineate the cooling window and the region of work extraction during the feedback stage, respectively, are however clearly not iso-discordant. To see this, just note that the condition for cooling $ \sin{\varphi} > \epsilon_\mathsf{S}/\epsilon_\mathsf{A} $ and the formula for the critical angle $ \varphi_\text{crit} $ in Equation~(\ref{eq:varphi_work_extr}) depend explicitly on $ \epsilon_\mathsf{A} $. Hence, as was the case for entanglement, one cannot conclude that the build-up of a certain amount of quantum correlations unambiguously heralds the transition into these important regimes of operation.

At most, one can try to find the \textit{minimum} quantumness of correlations required for effective cooling. The discord $ \delta_\text{min} $ at the limiting setup $ \sin{\varphi} = \epsilon_\mathsf{S} $ can be easily found from Equation~\eqref{eq:discord_m}. Then, one can claim that cooling is guaranteed in all protocols achieving $ \delta(\hat{\varrho}_\text{m}) > \delta_\text{min} $. In the case depicted in Figure~\ref{fig:correlations}c (\textit{i.e}.,~$ \epsilon_\mathsf{S} = 0.4 $), $ \delta_\text{{min}}\simeq 1.35\times 10^{-2} $ nats, which roughly corresponds to the first contour line.


\section{Conclusions}\label{sec:conclusions}

In this paper, we have analysed a simple, open-system entropy-reduction algorithm from two~complementary viewpoints. On the one hand, we have evaluated the energy changes throughout the process, identifying the input work $ \mathcal{W} $, the entropy reduction per cycle $ \mathcal{P} $ and the residual heating of the environment $ \mathcal{Q} $. Equipped with these thermodynamic variables, we have looked into the performance optimization of the protocol, borrowing tools from thermal engineering. Specifically,~we have introduced the coefficient of performance $ \varepsilon = \mathcal{P}/\mathcal{W} $ in order to assess the energy efficiency of the cycle and an alternative figure of merit $ \chi = \varepsilon\mathcal{P} $ that balances the energy efficiency and the effectiveness of the process. The maximization of either of these objective functions allows one to identify thermodynamically optimal working points in the space of parameters of the cycle.

On the other hand, we have studied the correlations built up between the system of interest and the external controller. In particular, we wanted to elucidate the connections linking those correlations to the performance optimization of the algorithm. Interestingly, we have found that for every choice of the measurement basis in which the controller interrogates the system, the quantum share of correlations, as measured by quantum discord, remains constant regardless of the entropy reduction rate $ \mathcal{P} $, since it does not depend on the polarization bias of the ancillary spins $\epsilon_\mathsf{A}$. Likewise, discord increases monotonically as the performance of the cycle is optimized in terms of the measurement direction. In contrast, neither the total correlations, nor their classical share relate so neatly with the figures of merit $ \varepsilon $ or $ \chi $, and specifically, they do not attain their global maxima at thermodynamically-optimal working points. Exactly the same can be said about quantum entanglement, which is another (somewhat more stringent) quantifier of the quantumness of correlations. We have also obtained the minimum amount of discord between the system and controller necessary for effectively cooling the~system.

This intriguing connection between thermodynamic performance and quantum correlations beyond entanglement poses the question of whether or not the latter act conclusively as resources in simple cooling protocols, such as the one considered here, or, more generally, in measurement-based quantum feedback control tasks. This open question certainly deserves detailed examination: it is not only relevant for practical applications of control theory in the rapidly developing field of quantum technologies, but also, from a more fundamental perspective, it may help to further clarify the role of quantum effects in thermodynamics and, broadly speaking, in the regulation of complex phenomena~\cite{Towards}.

\vspace{12pt}
\acknowledgments{\textbf{Acknowledgements}: We gratefully acknowledge financial support from the European Cooperation in Science and Technology (COST) Action MP1209 ``Thermodynamics in the quantum regime''.
Pietro Liuzzo-Scorpo acknowledges financial support from the University of Nottingham (Graduate School Travel Prize 2015).
 Luis A. Correa acknowledges financial support from Spanish Ministry of Economy and Competitiveness (MINECO)  (Project No.~FIS2013-40627-P) and Generalitat de Catalunya Consejo Interdepartamental de Investigaci\'on e Innovaci\'on Tecnol\'ogica (CIRIT 
 Project No.~2014 SGR 966). Rebecca Schmidt acknowledges financial support
 by the Academy of Finland through its Centres of Excellence Programme (2015-2017) under project number 284621.
Rebecca Schmidt and Gerardo Adesso acknowledge financial support from the Foundational Questions Institute (Project No.~FQXi-RFP3-1317).
Gerardo Adesso acknowledges financial support from the European Research Council (ERC Starting Grant Agreement No.~637352).
We warmly thank Davide Girolami and Isabela A. Silva for fruitful~discussions.}

{\textbf{Author Contributions:} Pietro Liuzzo-Scorpo and Luis A. Correa performed the main theoretical analysis, with inputs from Rebecca Schmidt and Gerardo Adesso. All authors contributed substantially to the discussions, development and writing of the paper. All authors read and approved the final mansucript.}


\appendix
\section*{\noindent Appendix: Explicit Formula for the Quantum Mutual Information}
\vspace{6pt}

The analytical expression of the mutual information after the (pre-)measurement reads:
	\begin{align}
	I(\hat{\varrho}_m)=&-\frac{1}{2}\log \left(\frac{1-\epsilon_\mathsf{S} \epsilon_\mathsf{A} \cos \varphi
	}{4}\right)-\frac{1}{2}\log (\epsilon_\mathsf{S} \epsilon_\mathsf{A} \cos \varphi+1)- \epsilon_\mathsf{S} \epsilon_\mathsf{A} \cos \varphi
	\arctanh(\epsilon_\mathsf{S} \epsilon_\mathsf{A} \cos \varphi)-\notag\\
	&-\frac{1}{2} \log \left(\frac{1-\epsilon_\mathsf{S} \cos \varphi }{4}\right)-\frac{1}{2}\log (\epsilon_\mathsf{S} \cos \varphi+1)- \epsilon_\mathsf{S} \cos
	\varphi \arctanh(\epsilon_\mathsf{S} \cos \varphi )-\notag\\
	&-\frac{1}{4}(1-\epsilon_\mathsf{S} ) (\epsilon_\mathsf{A} -1) \log
	\left(\frac{1}{4} (\epsilon_\mathsf{S} -1) (\epsilon_\mathsf{A} -1)\right)-\notag\\
	&-(\epsilon_\mathsf{S} +1) (\epsilon_\mathsf{A} -1) \log
	\left(-\frac{1}{4} (\epsilon_\mathsf{S} +1) (\epsilon_\mathsf{A} -1)\right)-\notag\\
	&-\frac{1}{4}(\epsilon_\mathsf{S} -1) (\epsilon_\mathsf{A} +1) \log
	\left(-\frac{1}{4} (\epsilon_\mathsf{S} -1) (\epsilon_\mathsf{A} +1)\right)+\notag\\
	&+(\epsilon_\mathsf{S} +1) (\epsilon_\mathsf{A} +1) \log
	\left(\frac{1}{4} (\epsilon_\mathsf{S} +1) (\epsilon_\mathsf{A} +1)\right)\notag~.
	\end{align}







\renewcommand\bibname{References}


%



\begin{thebibliography}{999}
		\providecommand{\natexlab}[1]{#1}
	
\bibitem[Anglin and Ketterle(2002)]{anglin2002bose}
	Anglin, J.R.; Ketterle, W.
	\newblock Bose--Einstein condensation of atomic gases.
	\newblock {\it Nature} {\bf 2002}, {\it 416},~211--218.
	
	\bibitem[Phillips(1998)]{phillips1998laser}
	Phillips, W.D.
	\newblock {Nobel Lecture: Laser Cooling and Trapping of Neutral Atoms.}
	\newblock {\it Rev. Mod. Phys.} {\bf 1998}, {\it 70},~721, doi:10.1103/RevModPhys.70.721.
	
	\bibitem[Masuhara \it{et~al.}(1988)Masuhara, Doyle, Sandberg, Kleppner,
	Greytak, Hess, and Kochanski]{masuhara1988evaporative}
	Masuhara, N.; Doyle, J.M.; Sandberg, J.C.; Kleppner, D.; Greytak, T.J.; Hess,
	H.F.; Kochanski, G.P.
	\newblock Evaporative cooling of spin-polarized atomic hydrogen.
	\newblock {\it Phys. Rev. Lett.} {\bf 1988}, {\it 61},~935--938.
	
	\bibitem[Hopkins \it{et~al.}(2003)Hopkins, Jacobs, Habib, and
	Schwab]{hopkins2003feedback}
	Hopkins, A.; Jacobs, K.; Habib, S.; Schwab, K.
	\newblock Feedback cooling of a nanomechanical resonator.
	\linebreak \newblock {\it Phys. Rev.~B} {\bf 2003}, {\it 68},~235328.
	
	\bibitem[Kleckner and Bouwmeester(2006)]{kleckner2006sub}
	Kleckner, D.; Bouwmeester, D.
	\newblock Sub-kelvin optical cooling of a micromechanical resonator.
	\newblock {\it Nature} {\bf 2006}, {\it 444},~75--78.
	
	\bibitem[Poggio \it{et~al.}(2007)Poggio, Degen, Mamin, and
	Rugar]{poggio2007feedback}
	Poggio, M.; Degen, C.; Mamin, H.; Rugar, D.
	\newblock Feedback cooling of a cantilever's fundamental mode below 5 mK.
	\newblock {\it Phys. Rev. Lett.} {\bf 2007}, {\it 99},~017201.
	
	\bibitem[Kosloff and Levy(2014)]{1310.0683v1}
	Kosloff, R.; Levy, A.
	\newblock Quantum Heat Engines and Refrigerators: Continuous Devices.
	\newblock {\it Annu. Rev. Phys. Chem.} {\bf 2014}, {\it 65},~365--393.
	
	\bibitem[Gelbwaser-Klimovsky \it{et~al.}(2015)Gelbwaser-Klimovsky, Niedenzu,
	and Kurizki]{gelbwaser2015thermodynamics}
	Gelbwaser-Klimovsky, D.; Niedenzu, W.; Kurizki, G.
	\newblock {Thermodynamics of Quantum Systems Under Dynamical Control.}
	\newblock {\it Adv. At. Mol. Opt. Phys.} {\bf 2015},
	{\it 64},~329--407.
	
	\bibitem[Kosloff(2013)]{e15062100}
	Kosloff, R.
	\newblock Quantum Thermodynamics: A Dynamical Viewpoint.
	\newblock {\it Entropy} {\bf 2013}, {\it 15},~2100--2128.
	
	\bibitem[Koski(2015)]{Koski2015}
	Koski, J.V.; Kutvonen, A.; Khaymovich, I.M.; Ala-Nissila, T.; Pekola, J.P.
	\newblock On-chip Maxwell's demon as an information-powered refrigerator.
	\newblock \textit{Phys. Rev. Lett}. \textbf{2015}, \textit{115}, 260602.
	
	\bibitem[Kutvonen(2015)]{Kutvonen2015}
	 Kutvonen, A.; Koski, J.; Ala-Nissila, T.
	\newblock Thermodynamics and efficiency of an autonomous on-chip Maxwell's demon.
	\newblock {\bf 2015}, arXiv:1509.08288.
	

	\bibitem[Palao \it{et~al.}(2001)Palao, Kosloff, and Gordon]{PhysRevE.64.056130}
	Palao, J.P.; Kosloff, R.; Gordon, J.M.
	\newblock Quantum thermodynamic cooling cycle.
	\newblock {\it Phys. Rev. E} {\bf 2001}, {\it 64},~056130.
	
	\bibitem[Gelbwaser-Klimovsky and Kurizki(2014)]{gelbwaser2014heat}
	Gelbwaser-Klimovsky, D.; Kurizki, G.
	\newblock Heat-machine control by quantum-state preparation: From quantum
	engines to refrigerators.
	\newblock {\it Phys. Rev. E} {\bf 2014}, {\it 90},~022102.
	
	\bibitem[Correa(2014)]{PhysRevE.89.042128}
	Correa, L.A.
	\newblock Multistage quantum absorption heat pumps.
	\newblock {\it Phys. Rev. E} {\bf 2014}, {\it 89},~042128.
	
	\bibitem[Rezek \it{et~al.}(2009)Rezek, Salamon, Hoffmann, and
	Kosloff]{rezek2009quantum}
	Rezek, Y.; Salamon, P.; Hoffmann, K.H.; Kosloff, R.
	\newblock The quantum refrigerator: The quest for absolute zero.
	\newblock {\it Europhys. Lett.} {\bf 2009}, {\it 85},~30008.
	
	\bibitem[Kol\'a\ifmmode~\check{r}\else \v{r}\fi{}
	\it{et~al.}(2012)Kol\'a\ifmmode~\check{r}\else \v{r}\fi{},
	Gelbwaser-Klimovsky, Alicki, and Kurizki]{PhysRevLett.109.090601}
	Kol\'a\ifmmode~\check{r}\else \v{r}\fi{}, M.; Gelbwaser-Klimovsky, D.; Alicki,
	R.; Kurizki, G.
	\newblock Quantum Bath Refrigeration towards Absolute Zero: Challenging the
	Unattainability Principle.
	\newblock {\it Phys. Rev. Lett.} {\bf 2012}, {\it 109},~090601.
	
	\bibitem[Levy \it{et~al.}(2012)Levy, Alicki, and Kosloff]{PhysRevE.85.061126}
	Levy, A.; Alicki, R.; Kosloff, R.
	\newblock Quantum refrigerators and the third law of thermodynamics.
	\newblock {\it Phys. Rev. E} {\bf 2012}, {\it 85},~061126.
	
	\bibitem[Allahverdyan \it{et~al.}(2010)Allahverdyan, Hovhannisyan, and
	Mahler]{PhysRevE.81.051129}
	Allahverdyan, A.E.; Hovhannisyan, K.; Mahler, G.
	\newblock Optimal refrigerator.
	\newblock {\it Phys. Rev. E} {\bf 2010}, {\it 81},~051129.
	
	\bibitem[Correa \it{et~al.}(2013)Correa, Palao, Adesso, and Alonso]{Correa2013}
	Correa, L.A.; Palao, J.P.; Adesso, G.; Alonso, D.
	\newblock Performance bound for quantum absorption refrigerators.
	\newblock {\it Phys. Rev. E} {\bf 2013}, {\it 87},~042131.
	
	\bibitem[Correa \it{et~al.}(2014)Correa, Palao, Adesso, and
	Alonso]{PhysRevE.90.062124}
	Correa, L.A.; Palao, J.P.; Adesso, G.; Alonso, D.
	\newblock Optimal performance of endoreversible quantum refrigerators.
	\newblock {\it Phys. Rev. E} {\bf 2014}, {\it 90},~062124.
	
	\bibitem[Kosloff and Feldmann(2010)]{kosloff2010optimal}
	Kosloff, R.; Feldmann, T.
	\newblock Optimal performance of reciprocating demagnetization quantum
	refrigerators.
	\newblock {\it Phys. Rev. E} {\bf 2010}, {\it 82},~011134.
	
	\bibitem[Correa \it{et~al.}(2015)Correa, Palao, and Alonso]{correa2015internal}
	Correa, L.A.; Palao, J.P.; Alonso, D.
	\newblock Internal dissipation and heat leaks in quantum thermodynamic cycles.
	\newblock {\it Phys. Rev. E} {\bf 2015}, {\it 92},~032136.
	
	\bibitem[Feldmann and Kosloff(2006)]{PhysRevE.73.025107}
	Feldmann, T.; Kosloff, R.
	\newblock Quantum lubrication: Suppression of friction in a first-principles
	four-stroke heat engine.
	\newblock {\it Phys. Rev. E} {\bf 2006}, {\it 73},~025107.
	
	\bibitem[Chen and Li(2012)]{0295-5075_97_4_40003}
	Chen, Y.-X.; Li, S.-W.
	\newblock Quantum refrigerator driven by current noise.
	\newblock {\it Europhys. Lett.} {\bf 2012}, {\it 97},~40003.
	
	\bibitem[Venturelli \it{et~al.}(2013)Venturelli, Fazio, and
	Giovannetti]{PhysRevLett.110.256801}
	Venturelli, D.; Fazio, R.; Giovannetti, V.
	\newblock Minimal Self-Contained Quantum Refrigeration Machine Based on Four
	Quantum Dots.
	\newblock {\it Phys. Rev. Lett.} {\bf 2013}, {\it 110},~256801.
	
	\bibitem[Belthangady \it{et~al.}(2013)Belthangady, Bar-Gill, Pham, Arai,
	Le~Sage, Cappellaro, and Walsworth]{PhysRevLett.110.157601}
	Belthangady, C.; Bar-Gill, N.; Pham, L.M.; Arai, K.; Le Sage, D.; Cappellaro,
	P.; Walsworth, R.L.
	\newblock Dressed-State Resonant Coupling between Bright and Dark Spins in
	Diamond.
	\newblock {\it Phys. Rev. Lett.} {\bf 2013}, {\it 110},~157601.
	
	\bibitem[Gelbwaser-Klimovsky \it{et~al.}(2015)Gelbwaser-Klimovsky,
	Szczygielski, Vogl, Sa\ss{}, Alicki, Kurizki, and Weitz]{PhysRevA.91.023431}
	Gelbwaser-Klimovsky, D.; Szczygielski, K.; Vogl, U.; Sa\ss{}, A.; Alicki, R.;
	Kurizki, G.; Weitz, M.
	\newblock Laser-induced cooling of broadband heat reservoirs.
	\newblock {\it Phys. Rev. A} {\bf 2015}, {\it 91},~023431.
	
	\bibitem[Steck \it{et~al.}(2004)Steck, Jacobs, Mabuchi, Bhattacharya, and
	Habib]{steck2004quantum}
	Steck, D.A.; Jacobs, K.; Mabuchi, H.; Bhattacharya, T.; Habib, S.
	\newblock Quantum feedback control of atomic motion in an optical cavity.
	\newblock {\it Phys. Rev. Lett.} {\bf 2004}, {\it 92},~223004.
	
	\bibitem[Bushev \it{et~al.}(2006)Bushev, Rotter, Wilson, Dubin, Becher,
	Eschner, Blatt, Steixner, Rabl, and Zoller]{bushev2006feedback}
	Bushev, P.; Rotter, D.; Wilson, A.; Dubin, F.; Becher, C.; Eschner, J.; Blatt,
	R.; Steixner, V.; Rabl, P.; Zoller, P.
	\newblock Feedback cooling of a single trapped ion.
	\newblock {\it Phys. Rev. Lett.} {\bf 2006}, {\it 96},~043003.
	
	\bibitem[Abah and Lutz(2014)]{1303.6558v1}
	Abah, O.; Lutz, E.
	\newblock Efficiency of heat engines coupled to nonequilibrium reservoirs.
	\newblock {\it Europhys. Lett.} {\bf 2014}, {\it 106},~20001.
	
	\bibitem[Correa \it{et~al.}(2014)Correa, Palao, Alonso, and Adesso]{Correa2014}
	Correa, L.A.; Palao, J.P.; Alonso, D.; Adesso, G.
	\newblock Quantum-enhanced absorption refrigerators.
	\newblock {\it Sci. Rep.} {\bf 2014}, {\it 4}, 3949, doi:10.1038/srep03949.
	
	\bibitem[Ro\ss{}nagel \it{et~al.}(2014)Ro\ss{}nagel, Abah, Schmidt-Kaler,
	Singer, and Lutz]{PhysRevLett.112.030602}
	Ro\ss{}nagel, J.; Abah, O.; Schmidt-Kaler, F.; Singer, K.; Lutz, E.
	\newblock Nanoscale Heat Engine Beyond the Carnot Limit.
	\newblock {\it Phys. Rev. Lett.} {\bf 2014}, {\it 112},~030602.
	
	\bibitem[Alicki and Gelbwaser-Klimovsky(2015)]{alicki2015non}
	Alicki, R.; Gelbwaser-Klimovsky, D.
	\newblock Non-equilibrium quantum heat machines.
	\newblock {\it New J. Phys. } {\bf 2015}, \textit{17},~115012.

	\bibitem[Niedenzu \it{et~al.}(2015)Niedenzu, Gelbwaser-Klimovsky, and Kurizki]{niedenzu}
	Niedenzu, W.; Gelbwaser-Klimovsky, D.; Kurizki, G.
	\newblock Performance limits of multilevel and multipartite quantum heat machines.
	\newblock {\it Phys. Rev. E} {\bf 2015}, {\it 92},~042123.
	
	\bibitem[Uzdin \it{et~al.}(2015)Uzdin, Levy, and Kosloff]{PhysRevX.5.031044}
	Uzdin, R.; Levy, A.; Kosloff, R.
	\newblock Equivalence of Quantum Heat Machines, and Quantum-Thermodynamic
	Signatures.
	\newblock {\it Phys. Rev. X} {\bf 2015}, {\it 5},~031044.
	
	\bibitem[Alicki(1979)]{alicki1979engine}
	Alicki, R.
	\newblock {The quantum open system as a model of the heat engine.}
	\newblock {\it J. Phys. A} {\bf 1979}, {\it 12},~L103, doi:10.1088/\linebreak 0305-4470/12/5/007.
	
	\bibitem[Kosloff(1984)]{kosloff1984quantum}
	Kosloff, R.
	\newblock A quantum mechanical open system as a model of a heat engine.
	\newblock {\it J. Chem. Phys.} {\bf 1984}, {\it
		80},~1625--1631.
	
	\bibitem[Boykin \it{et~al.}(2002)Boykin, Mor, Roychowdhury, Vatan, and
	Vrijen]{boykin2002algorithmic}
	Boykin, P.O.; Mor, T.; Roychowdhury, V.; Vatan, F.; Vrijen, R.
	\newblock Algorithmic cooling and scalable NMR quantum computers.
	\newblock {\it Proc. Natl. Acad. Sci. USA} {\bf 2002},
	{\it 99},~3388--3393.
	
	\bibitem[Fernandez \it{et~al.}(2004)Fernandez, Lloyd, Mor, and
	Roychowdhury]{fernandez2004algorithmic}
	Fernandez, J.M.; Lloyd, S.; Mor, T.; Roychowdhury, V.
	\newblock Algorithmic cooling of spins: A practicable method for increasing
	polarization.
	\newblock {\it Int. J. Quantum Inf.} {\bf 2004}, {\it
		2},~461--477.
	
	\bibitem[Baugh \it{et~al.}(2005)Baugh, Moussa, Ryan, Nayak, and
	Laflamme]{baugh2005experimental}
	Baugh, J.; Moussa, O.; Ryan, C.A.; Nayak, A.; Laflamme, R.
	\newblock Experimental implementation of heat-bath algorithmic cooling using
	solid-state nuclear magnetic resonance.
	\newblock {\it Nature} {\bf 2005}, {\it 438},~470--473.
	
	\bibitem[Ryan \it{et~al.}(2008)Ryan, Moussa, Baugh, and Laflamme]{ryan2008spin}
	Ryan, C.; Moussa, O.; Baugh, J.; Laflamme, R.
	\newblock Spin based heat engine: Demonstration of multiple rounds of
	algorithmic cooling.
	\newblock {\it Phys. Rev. Lett.} {\bf 2008}, {\it 100},~140501.
	
	\bibitem[Lloyd(2000)]{Lloyd2000}
	Lloyd, S.
	\newblock Coherent quantum feedback.
	\newblock {\it Phys. Rev. A} {\bf 2000}, {\it 62},~022108.
	
	\bibitem[Habib \it{et~al.}(2002)Habib, Jacobs, and Mabuchi]{habib2002quantum}
	Habib, S.; Jacobs, K.; Mabuchi, H.
	\newblock {Quantum Feedback Control.}
	\newblock {\it Los Alamos Sci.} {\bf 2002}, {\it 27}, 126--135.


	\bibitem[Ollivier and Zurek(2001)]{olliver20011}
	Ollivier, H.; Zurek, W.H.
	\newblock Quantum Discord: A Measure of the Quantumness of Correlations.
	\newblock {\it Phys. Rev.~Lett.} {\bf 2002}, {\it 88},~017901.
	
	\bibitem[Henderson and Vedral(2001)]{henderson20011}
	Henderson, L.; Vedral, V.
	\newblock Classical, quantum and total correlations.
	\newblock {\it J. Phys. A} {\bf 2001}, {\it 34},~6899-6905.
	
	\bibitem[Parrondo \it{et~al.}(2015)Parrondo, Horowitz, and
	Sagawa]{parrondo2015thermodynamics}
	Parrondo, J.M.R.; Horowitz, J.M.; Sagawa, T.
	\newblock Thermodynamics of information.
	\newblock {\it Nat. Phys.} {\bf 2015}, {\it 11},~131--139.

	\bibitem[Sagawa and Ueda(2008)]{sagawa2008second}
	Sagawa, T.; Ueda, M.
	\newblock Second law of thermodynamics with discrete quantum feedback control.
	\linebreak \newblock {\it Phys. Rev.~Lett.} {\bf 2008}, {\it 100},~080403.
	
	\bibitem[Park \it{et~al.}(2013)Park, Kim, Sagawa, and Kim]{park2013heat}
	Park, J.J.; Kim, K.-H.; Sagawa, T.; Kim, S.W.
	\newblock Heat engine driven by purely quantum information.
	\linebreak \newblock {\it Phys. Rev. Lett.} {\bf 2013}, {\it 111},~230402.
	

	\bibitem{Dong2010}
	Dong, D.; Petersen, I.R.
	\newblock Quantum control theory and applications: A survey.
	\newblock {\it IET Control Theory Appl.} {\bf 2010}, {\it 4},~2651--2671.

	\bibitem{Wise2010}
	{Wiseman, H.M.; Milburn, G.J.}
	\newblock \textit{Quantum Measurement and Control};
	\newblock Cambridge University Press: Cambridge, UK, 2010.

	\bibitem[Doherty \it{et~al.}(2000)Doherty, Habib, Jacobs, Mabuchi, and
	Tan]{PhysRevA.62.012105}
	Doherty, A.C.; Habib, S.; Jacobs, K.; Mabuchi, H.; Tan, S.M.
	\newblock Quantum feedback control and classical control theory.
	\newblock {\it Phys. Rev. A} {\bf 2000}, {\it 62},~012105.

	\bibitem[Touchette(2004)]{Touch2004}
	Touchette, H.; Lloyd, S.
	\newblock Information-theoretic approach to the study of control systems.
	\newblock {\it Physica A} {\bf 2004}, {\it 331},~140--172.	
	
	\bibitem{Yama2014}
	Yamamoto, N.
	\newblock Coherent versus measurement feedback: Linear systems theory for quantum information.
	\newblock {\it Phys. Rev. X} {\bf 2014}, {\it 4},~041029.


	\bibitem{Wise1993}
	Wiseman, H.M.; Milburn, G.J.
	\newblock Quantum theory of optical feedback via homodyne detection.
	\newblock {\it Phys. Rev.~Lett.} {\bf 1993}, {\it 70},~548--551.

	\bibitem{Gough2009}
	Gough, J.E.; Wildfeuer, S.
	\newblock Enhancement of field squeezing using coherent feedback.
	\newblock {\it Phys. Rev. A} {\bf 2009}, {\it 80},~042107.

	\bibitem[Horowitz and Jacobs(2014)]{PhysRevE.89.042134}
	Horowitz, J.M.; Jacobs, K.
	\newblock Quantum effects improve the energy efficiency of feedback control.
	\newblock {\it Phys. Rev. E} {\bf 2014}, {\it 89},~042134.
	
	\bibitem[Allahverdyan \it{et~al.}(2004)Allahverdyan, Balian, and
	Nieuwenhuizen]{allahverdyan2004maximal}
	Allahverdyan, A.E.; Balian, R.; Nieuwenhuizen, T.M.
	\newblock Maximal work extraction from finite quantum systems.
	\newblock {\it Europhys. Lett.} {\bf 2004}, {\it 67},~565--571.
	
	\bibitem[Gordon and Ng(2000)]{Gordon2000}
	Gordon, J.M.; Ng, K.C.
	\newblock {\it Cool Thermodynamics}; Cambridge International Science
	Publishing: Cambridge, UK, 2000.

	\bibitem[Gordon(1991)]{gordon1991generalized}
	Gordon, J.M.
	\newblock {Generalized power versus efficiency characteristics of heat engines:
	The thermoelectric generator as an instructive illustration.}
	\newblock {\it Am. J. Phys.} {\bf 1991}, {\it 59},~551--555.
	
	\bibitem[Hoffmann \it{et~al.}(1997)Hoffmann, Burzler, and
	Schubert]{hoffmann1997endoreversible}
	Hoffmann, K.H.; Burzler, J.M.; Schubert, S.
	\newblock {Endoreversible thermodynamics.}
	\newblock {\it J. Non-Equilib. Thermodyn.} {\bf 1997}, {\it 22},~311--355.
	
	\bibitem[Yan and Chen(1990)]{0022-3727_23_2_002}
	Yan, Z.; Chen, J.
	\newblock {A class of irreversible Carnot refrigeration cycles with a general
	heat transfer law.}
	\newblock {\it J. Phys. D} {\bf 1990}, {\it 23}, doi:10.1088/0022-3727/23/2/002.
	
	\bibitem[de~Tom\'as \it{et~al.}(2012)de~Tom\'as, Hern\'andez, and
	Roco]{PhysRevE.85.010104}
	De~Tom\'as, C.; Hern\'andez, A.C.; Roco, J.M.M.
	\newblock Optimal low symmetric dissipation Carnot engines and refrigerators.
	\newblock {\it Phys. Rev. E} {\bf 2012}, {\it 85},~010104.
	
	\bibitem[Wang \it{et~al.}(2012)Wang, Li, Tu, Hern\'andez, and
	Roco]{PhysRevE.86.011127}
	Wang, Y.; Li, M.; Tu, Z.C.; Hern\'andez, A.C.; Roco, J.M.M.
	\newblock Coefficient of performance at maximum figure of merit and its bounds
	for low-dissipation Carnot-like refrigerators.
	\newblock {\it Phys. Rev. E} {\bf 2012}, {\it 86},~011127.

	\bibitem[Horodecki \it{et~al.}(2009)Horodecki, Horodecki, Horodecki, and
	Horodecki]{RevModPhys.81.865}
	Horodecki, R.; Horodecki, P.; Horodecki, M.; Horodecki, K.
	\newblock Quantum entanglement.
	\newblock {\it Rev. Mod. Phys.} {\bf 2009}, {\it 81},~865--942.
	
	\bibitem[Schr{\"o}dinger(1935)]{schrodinger1935gegenwartige}
	Schr{\"o}dinger, E.
	\newblock Die gegenw{\"a}rtige Situation in der Quantenmechanik.
	\newblock {\it Naturwissenschaften} {\bf 1935}, {\it 23},~823--828. (In German)
	
	\bibitem[Einstein \it{et~al.}(1935)Einstein, Podolsky, and
	Rosen]{einstein1935can}
	{Einstein, A.; Podolsky, B.; Rosen, N.}
	\newblock Can quantum-mechanical description of physical reality be considered
	complete?
	\newblock {\it Phys. Rev.} {\bf 1935}, {\it 47},~777, doi:10.1103/PhysRev.47.777.
	
	\bibitem[Bell(1964)]{bell1964einstein}
	Bell, J.S.
	\newblock On the Einstein Podolsky Rosen paradox.
	\newblock {\it Physics} {\bf 1964}, {\it 1},~195--200.
	
	\bibitem[Bennett \it{et~al.}(1993)Bennett, Brassard, Cr\'epeau, Jozsa, Peres,
	and Wootters]{PhysRevLett.70.1895}
	Bennett, C.H.; Brassard, G.; Cr\'epeau, C.; Jozsa, R.; Peres, A.; Wootters,
	W.K.
	\newblock Teleporting an unknown quantum state via dual classical and
	Einstein--Podolsky--Rosen channels.
	\newblock {\it Phys. Rev. Lett.} {\bf 1993}, {\it 70},~1895--1899.
	
	\bibitem[Bouwmeester \it{et~al.}(1997)Bouwmeester, Pan, Mattle, Eibl,
	Weinfurter, and Zeilinger]{bouwmeester1997experimental}
	Bouwmeester, D.; Pan, J.-W.; Mattle, K.; Eibl, M.; Weinfurter, H.; Zeilinger, A.
	\newblock Experimental quantum teleportation.
	\newblock {\it Nature} {\bf 1997}, {\it 390},~575--579.
	
	\bibitem[Bennett and Wiesner(1992)]{bennett19921}
	Bennett, C.H.; Wiesner, S.J.
	\newblock Communication via one- and two-particle operators on
	Einstein--Podolsky--Rosen states.
	\newblock {\it Phys. Rev. Lett.} {\bf 1992}, {\it 69},~2881--2884.
	
	\bibitem[Mattle \it{et~al.}(1996)Mattle, Weinfurter, Kwiat, and
	Zeilinger]{PhysRevLett.76.4656}
	{Mattle, K.; Weinfurter, H.; Kwiat, P.G.; Zeilinger, A.}
	\newblock Dense Coding in Experimental Quantum Communication.
	\newblock {\it Phys. Rev. Lett.} {\bf 1996}, {\it 76},~4656--4659.
	
	\bibitem[Ekert(1991)]{ekert19911}
	Ekert, A.K.
	\newblock Quantum cryptography based on Bell's theorem.
	\newblock {\it Phys. Rev. Lett.} {\bf 1991}, {\it 67},~661--663.
	
	\bibitem[Jennewein \it{et~al.}(2000)Jennewein, Simon, Weihs, Weinfurter, and
	Zeilinger]{PhysRevLett.84.4729}
	Jennewein, T.; Simon, C.; Weihs, G.; Weinfurter, H.; Zeilinger, A.
	\newblock Quantum Cryptography with Entangled Photons.
	\newblock {\it Phys. Rev. Lett.} {\bf 2000}, {\it 84},~4729--4732.
	
	\bibitem[Huelga \it{et~al.}(1997)Huelga, Macchiavello, Pellizzari, Ekert,
	Plenio, and Cirac]{PhysRevLett.79.3865}
	{Huelga, S.F.; Macchiavello, C.; Pellizzari, T.; Ekert, A.K.; Plenio, M.B.;
	Cirac, J.I.}
	\newblock Improvement of Frequency Standards with Quantum Entanglement.
	\newblock {\it Phys. Rev. Lett.} {\bf 1997}, {\it 79},~3865--3868.
	
	\bibitem[Wootters(1998)]{PhysRevLett.80.2245}
	{Wootters, W.K.}
	\newblock Entanglement of Formation of an Arbitrary State of Two Qubits.
	\newblock {\it Phys. Rev. Lett.} {\bf 1998}, {\it 80},~2245--2248.

	\bibitem[Hovhannisyan \it{et~al.}(2013)]{WorkEnt}
	{Hovhannisyan, K.V.; Perarnau-Llobet, M.; Huber, M.; Ac\'in, A.}
	\newblock Entanglement Generation is Not Necessary for Optimal Work Extraction.
	\newblock {\it Phys. Rev. Lett.} {\bf 2013}, {\it 111},~240401.
	
	\bibitem[Modi \it{et~al.}(2012)Modi, Brodutch, Cable, Paterek, and
	Vedral]{RevModPhys.84.1655}
	Modi, K.; Brodutch, A.; Cable, H.; Paterek, T.; Vedral, V.
	\newblock The classical-quantum boundary for correlations: Discord and related
	measures.
	\newblock {\it Rev. Mod. Phys.} {\bf 2012}, {\it 84},~1655--1707.
	
  \bibitem[Cavalcanti \it{et~al.}(2011)]{operd1}
  Cavalcanti, D.; Aolita, L.; Boixo, S.; Modi, K.; Piani, M.; Winter, A.
  \newblock  Operational interpretations of quantum discord.
  \newblock {\it Phys. Rev. A} {\bf 2011}, {\it 83},~032324.

  \bibitem[Madhok and Datta (2011)]{operd2}
  Madhok, V.; Datta, A.
  \newblock  Interpreting quantum discord through quantum state merging.
  \newblock {\it Phys. Rev. A} {\bf 2011}, {\it 83},~032323.

  \bibitem[Pirandola (2014)]{operdc}
  Pirandola, S. \newblock
  Quantum discord as a resource for quantum cryptography.
  \newblock {\it Sci. Rep.} {\bf 2014}, {\it 4},~6956, doi:10.1038/srep06956.

	\bibitem[Girolami \it{et~al.}(2013)Girolami, Tufarelli, and
	Adesso]{PhysRevLett.110.240402}
	Girolami, D.; Tufarelli, T.; Adesso, G.
	\newblock Characterizing Nonclassical Correlations via Local Quantum
	Uncertainty.
	\newblock {\it Phys. Rev. Lett.} {\bf 2013}, {\it 110},~240402.
	
	\bibitem[Girolami \it{et~al.}(2014)Girolami, Souza, Giovannetti, Tufarelli,
	Filgueiras, Sarthour, Soares-Pinto, Oliveira, and
	Adesso]{PhysRevLett.112.210401}
	Girolami, D.; Souza, A.M.; Giovannetti, V.; Tufarelli, T.; Filgueiras, J.G.;
	Sarthour, R.S.; Soares-Pinto, D.O.; Oliveira, I.S.; Adesso, G.
	\newblock Quantum Discord Determines the Interferometric Power of Quantum
	States.
	\linebreak \newblock {\it Phys. Rev. Lett.} {\bf 2014}, {\it 112},~210401.

  \bibitem[Zurek (2003)]{operdz}
  Zurek, W.H. \newblock
  Quantum discord and Maxwell's demons.
  \newblock {\it Phys. Rev. A} {\bf 2003}, {\it 67},~012320.

  \bibitem[Girolami, Schmidt, Adesso (2015)]{Towards}
  Girolami, D.; Schmidt, R.; Adesso, G.
  \newblock Towards quantum cybernetics.
  \newblock {\it Ann. Phys.} {\bf 2015}, {\it 527},~757--764.

\end{thebibliography}
\end{document}